\documentclass[twocolumn]{aastex7}
\usepackage[normalem]{ulem}
\usepackage{makecell}
\usepackage{graphics,xcolor}
\usepackage{amsmath}
\usepackage{soul}
\usepackage{CJK}
\usepackage{listings}
\usepackage{comment}

\newcommand{\lbolsunnum}{$3.828\times 10^{33}$\,erg/s}

\newcommand{\lbol}{L_{\rm bol}}
\newcommand{\fbol}{F_{\rm bol}}
\newcommand{\teff}{T_{\rm eff}}
\newcommand{\lsun}{L_{\sun}}
\newcommand{\gbp}{{G_\mathrm{BP}}}
\newcommand{\grp}{{G_\mathrm{RP}}}

\newcommand{\Msun}{\mbox{${\rm M_{\sun}}$}}
\newcommand{\Mjup}{\mbox{${\rm M_{\rm Jup}}$}}
\newcommand{\Rjup}{\mbox{${\rm R_{\rm Jup}}$}}
\newcommand{\Rsun}{\mbox{${\rm R_{\sun}}$}}

\begin{document}

\title{The Diversity of Cold Worlds: Age and Characterization of the Exoplanet COCONUTS-2b}
\shorttitle{Age and Characterization of COCONUTS-2b}
\shortauthors{Kiman et al.}
 
\accepted{22-Nov-2025}
\submitjournal{The Astronomical Journal}

\author[0000-0003-2102-3159]{Rocio Kiman}
\affil{Department of Astronomy, California Institute of Technology, Pasadena, CA 91125, USA}
\email{rociokiman@gmail.com}
\correspondingauthor{Rocio Kiman}

\author[0000-0002-5627-5471]{Charles A. Beichman}
\affil{Department of Astronomy, California Institute of Technology, Pasadena, CA 91125, USA}
\affil{Infrared Processing and Analysis Center, California Institute of Technology, 1200 E. California Boulevard, Pasadena, CA 91125, USA}
\email{}

\author[0000-0001-6733-4118]{Azul Ruiz Diaz}
\affil{Department of Astrophysics, American Museum of Natural History, Central Park West at 79th St, New York, NY 10024, USA}
\email{}

\author[0000-0001-6251-0573]{Jacqueline K. Faherty}
\affil{Department of Astrophysics, American Museum of Natural History, Central Park West at 79th St, New York, NY 10024, USA}
\email{}

\author[0000-0002-9420-4455]{Brianna Lacy}
\affiliation{Department of Astronomy and Astrophysics, University of California, Santa Cruz, CA, USA}
\email{}

\author[0000-0002-2011-4924]{Genaro Su\'arez}
\affil{Department of Astrophysics, American Museum of Natural History, Central Park West at 79th St, New York, NY 10024, USA}
\email{}

\author[0000-0001-7519-1700]{Federico Marocco}
\affiliation{Infrared Processing and Analysis Center, California Institute of Technology, 1200 E. California Boulevard, Pasadena, CA 91125, USA}
\email{}

\author[0000-0003-4269-260X]{J. Davy Kirkpatrick}
\affiliation{Infrared Processing and Analysis Center, California Institute of Technology, 1200 E. California Boulevard, Pasadena, CA 91125, USA}
\email{}

\author[0000-0002-2592-9612]{Jonathan Gagn\'e}
\affiliation{Plan\'etarium de Montr\'eal, Espace pour la Vie, 4801 av. Pierre-de Coubertin, Montr\'eal, Qu\'ebec, Canada}
\affiliation{Trottier Institute for Research on Exoplanets, Universit\'e de Montr\'eal, D\'epartement de Physique, C.P.~6128 Succ. Centre-ville, Montr\'eal, QC H3C~3J7, Canada}
\email{}

\author[0009-0005-7267-7760]{Jessica Copeland}
\affiliation{Department of Physics, Astronomy and Mathematics, University of Hertfordshire, Hatfield, UK}
\email{}

\author[0000-0003-4600-5627]{Ben Burningham}
\affiliation{Department of Physics, Astronomy and Mathematics, University of Hertfordshire, Hatfield, UK}
\email{}

\author[0000-0001-8818-1544]{Niall Whiteford}
\affiliation{Department of Astrophysics, American Museum of Natural History, Central Park West at 79th St, New York, NY 10024, USA}
\email{}

\author[0000-0003-4225-6314]{Melanie J. Rowland}
\affiliation{Department of Astronomy, University of Texas at Austin, Austin, TX, USA}
\email{}

\author[0000-0001-8170-7072]{Daniella C. Bardalez Gagliuffi}
\affiliation{Department of Physics \& Astronomy, Amherst College, Amherst, MA, USA}
\email{}

\author[0000-0003-0489-1528]{Johanna M. Vos}
\affiliation{School of Physics, Trinity College Dublin, The University of Dublin, Dublin 2, Ireland}
\affil{Department of Astrophysics, American Museum of Natural History, Central Park West at 79th St, New York, NY 10024, USA}
\email{}

\author[0000-0002-6294-5937]{Adam C. Schneider}
\affil{United States Naval Observatory, Flagstaff Station, 10391 West Naval Observatory Rd., Flagstaff, AZ 86005, USA}
\email{}

\author[0000-0003-4636-6676]{Eileen C. Gonzales}
\affil{Department of Physics and Astronomy, San Francisco State University, 1600 Holloway Ave., San Francisco, CA 94132, USA}
\email{}

\author[0000-0003-0548-0093]{Sherelyn Alejandro Merchan}
\affil{Department of Astrophysics, American Museum of Natural History, Central Park West at 79th St, New York, NY 10024, USA}
\affiliation{Department of Physics, Graduate Center, City University of New York, 365 5th Ave., New York, NY 10016, USA}
\email{}

\author[0000-0003-4083-9962]{Austin Rothermich}
\affil{Department of Astrophysics, American Museum of Natural History, Central Park West at 79th St, New York, NY 10024, USA}
\affiliation{Department of Physics, Graduate Center, City University of New York, 365 5th Ave., New York, NY 10016, USA}
\affiliation{Department of Physics and Astronomy, Hunter College, City University of New York, 695 Park Avenue, New York, NY, 10065, USA}
\affiliation{Backyard Worlds: Planet 9}
\email{}

\author[0000-0002-4424-4766]{Richard Smart}
\affiliation{Osservatorio Astronomico di Torino: Pino Torinese, Piemonte, IT}
\email{}

\author[0000-0003-4142-1082]{Edgardo Costa}
\affiliation{Departamento de Astronomia, Universidad de Chile, Casilla 36-D, Santiago, Chile}
\email{}

\author[0000-0003-1454-0596]{Rene A. Mendez}
\affiliation{Departamento de Astronomia, Universidad de Chile, Casilla 36-D, Santiago, Chile}
\email{}


\begin{abstract} 

Studying cold brown dwarfs is key to understanding the diverse characteristics of cold giant exoplanets atmospheres. 
COCONUTS-2, is a wide binary system composed of a T9 brown dwarf and an M3 star, which presents a unique opportunity to characterize a cold benchmark brown dwarf. 
As part of a JWST program to study the range of physical and atmospheric properties of the coldest brown dwarfs, we obtained NIRSpec G395H spectra (R$\sim2700$, $2.87-5.13~\,{\rm \mu m}$) and MIRI F1000W, F1280W, and F1800W photometry for COCONUTS-2b. 
In this work, we find a $99\%$ probability of the system belonging to the Corona of Ursa Major ($414\pm23$ Myr) using BANYAN $\Sigma$ and its full kinematics. We also re-estimate the astrometry of COCONUTS-2b using the MIRI data. We support the membership with a comparison of rotation period, metallicity and C/O ratio of the group with those of the COCONUTS-2 system. 
We also calculate its bolometric luminosity, which combined with our age estimation, allows us to derive its mass, effective temperature, surface gravity, and radius with high precision. 
As a result of our analysis, we support the conclusion that COCONUTS-2b is a planetary mass object ($7.5\pm0.4\,\Mjup$) which was likely formed via the same mechanism as stars. 
In addition we compare the JWST spectrum to another object in the sample, J082507.35+280548.5 (0825+2805), a Y0.5 brown dwarf, which is a candidate member of the same moving group, but has a lower mass (${3.7}\pm{0.2}\,\Mjup$). We identify absorption feature differences which could indicate that 0825+2805 has stronger vertical mixing.

\end{abstract}

\section{Introduction}
\label{sec:intro}

Brown dwarfs are objects with masses $<78.5\,\Mjup$ \citep[][with previous estimations such as $75\,\Mjup$ from \citealt{Kumar1963}, and $70\,\Mjup$ from \citealt{Dupuy2017}]{Chabrier2023} that, unlike stars, do not burn hydrogen in their core. Although they are more massive than exoplanets, their cold atmospheres resemble these objects. Furthermore, brown dwarfs can be observed independently of a host star, making them the perfect laboratory to study exoplanets atmospheres.
However, one of the main difficulties when studying brown dwarfs, is the luminosity-age-temperature degeneracy \citep[e.g.,][]{Burrows1997}. After formation, brown dwarfs cool with time, and go through different spectral types \citep[e.g.,][]{Saumon2008}. 
For example, if we measure a cold effective temperature for an object (for example $1900$\,K), we cannot distinguish if this object is high-mass and old ($75\,\Mjup$, $\sim 5$\,Gyr), or low-mass and young ($13\,\Mjup$, $\sim 10$\,Myr). Studying atmospheric composition of brown dwarfs is even more complex, given that to the problem described above, we also add parameters such as C/O ratio and the $K_{zz}$ diffusion parameter \citep{Saumon2012}. 
Benchmark objects with known properties that reduce the degrees of freedom are key to refining models and ultimately fully understanding their atmospheres \citep{Pinfield2006,Liu2008,Phillips2024}. 
Brown dwarfs in a binary with a main sequence star \citep[e.g.,][]{Crepp2018, Rickman2020, Rothermich2024, Phillips2024,Xuan2024}, or that belong to moving groups \citep[e.g.,][]{Aller2016, Gagne2023}, are perfect benchmark candidates given that we can use the primary, or group members, to estimate properties of the brown dwarf such as metallicity and age.

In this work we study COCONUTS-2b, or WISEP J075108.79-763449.6 \citep{Kirkpatrick2011}, a cold \citep[$430$\,K,][]{Zhang2021b} brown dwarf, which was one of the $12$ objects observed as part of the JWST GO 2124 program from Cycle 1 (PI J. Faherty). This program has the goal of explaining the spread in the Spitzer IRAC color-magnitude diagram of cold brown dwarfs, spanning $1-2$ absolute [4.5] magnitude. 
COCONUTS-2b is particularly interesting given that it is in a wide binary system with an M3 type star, separated by $6471$\,AU \citep{Zhang2021b}. Furthermore, COCONUTS-2 is a potential member of the young moving group Corona of Ursa Major \citep[CUMA,][]{Marocco2024}, making the system ideal to study the properties of cold brown dwarfs anchored on the properties of the primary and the group. 
COCONUTS-2b is also an under-luminous object in the Spitzer color-magnitude diagram, and also exhibits K-band flux suppression characteristic of T subdwarfs, as seen in a low-SNR spectrum \citep{Kirkpatrick2011,Kirkpatrick2021}, making it a compelling target for JWST spectroscopic follow-up.
The JWST GO 2124 program collected spectra and photometry to constrain key physical properties, including age, metallicity, clouds, C/O ratio, vertical mixing, chemical disequilibrium, elemental abundances and binarity. This type of analysis has been extensively applied to warmer objects such as L and early T dwarfs \citep{Faherty2012,Faherty2016,Burgasser2010b}, but remains scarce for late T and Y dwarfs \citep{Zhang2021}.

The goal of this work is to study the properties of the COCONUTS-2 system in the context of the moving group to which it has been tentatively linked. In Section~\ref{sec:data} we describe the JWST spectrum and photometry we obtained in Cycle 1, as well as the compilation of data available in the literature. 
In Section~\ref{sec:age} we analyze the membership of COCONUTS-2 to the Corona of Ursa Major, which we use to establish the age of the system. We also use the members of CUMA and all the information compiled on the COCONUTS-2 system to comment on its formation mechanism.
In Section~\ref{sec:lum} we combine all the available data to estimate the bolometric luminosity of both components of the system, and estimate mass, radius, effective temperature and surface gravity, assuming the age of CUMA. 
In Section~\ref{sec:atmosphere} we present a detailed analysis of the molecules found in the atmosphere of COCONUTS-2b by comparing the JWST spectrum to molecular opacity cross-sections, and a forward modeling analysis of all the available spectra for this object. 
In addition, we compare COCONUTS-2b to a second object from the JWST GO 2124 program, J082507.35+280548.5, which has high probability of belonging to the same cluster. Finally in Section~\ref{sec:conclusions} we include the conclusions of this work.

\section{Data}
\label{sec:data}

As part of JWST GO 2124 program, we collected near-infrared spectra with the highest attainable resolution (${\rm R}\sim 2700$), and mid-infrared photometry of COCONUTS-2b. In this section we describe the new data, and the compilation of available data from the literature for both components of the system.

\subsection{Spectra}
\label{subsec:spectra}

\subsubsection{JWST/NIRSpec Spectrum}
\label{subsubsection:spectrum_jwst}

\begin{figure*}[ht!]
\begin{center}
\includegraphics[width=\linewidth]{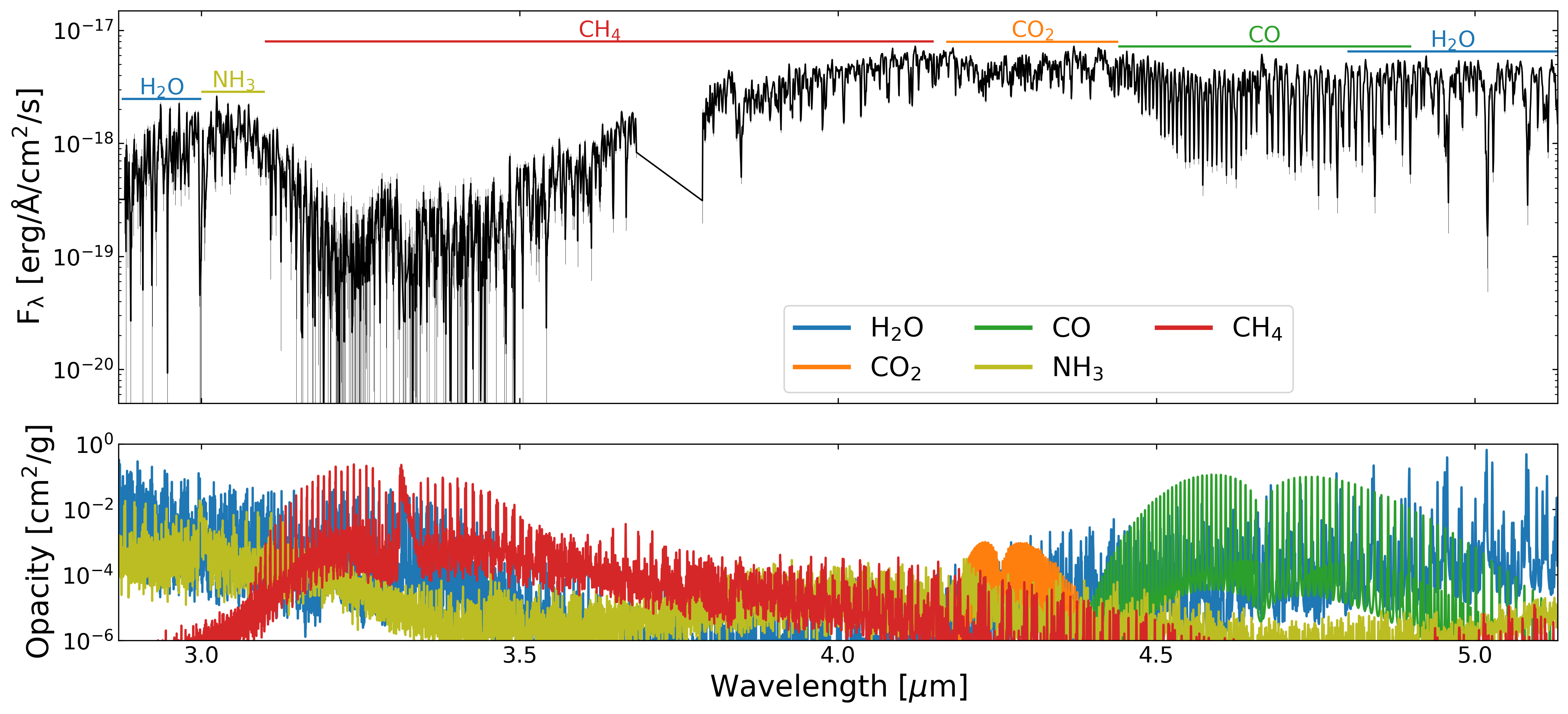}
\caption{Top panel shows the JWST spectrum of COCONUTS-2b in the filter G395H with horizontal lines indicating the molecular features which are recognizable in the spectrum. We also included the uncertainty in the flux in light-gray. Bottom panel shows the opacities for each identified molecule which we obtained from the DACE database \citep{Barber2006,Rothman2010,Yurchenko2011,Azzam2016,Hargreaves2020,Grimm2021,Tennyson2024}. See Section~\ref{sec:atmosphere} for a detailed description and discussion.} 
\label{fig:opacities}
\end{center}
\end{figure*}

As part of JWST Cycle 1, we obtained a spectrum of COCONUTS-2b using the JWST Near Infrared Spectrograph \citep[NIRSpec,][]{Jakobsen2022} G395H disperser, which provides the highest resolution spectrum attainable (average ${\rm R}\sim\,2700$) in the wavelength range $2.87-5.13~\,{\rm \mu m}$\footnote{The specific observations analyzed can be accessed via \dataset[https://doi.org/10.17909/rxm9-qd05]{https://doi.org/10.17909/rxm9-qd05}.}. 
We show the JWST spectrum of COCONUTS-2b in Figure~\ref{fig:opacities}. In this work, we used the data generated using the calibration software version 1.17.1., and the file that was made on 2025-03-19. This spectrum has a median SNR of 16 in the full wavelength range, or 30 in the range $>3.79~\mu$m.

\subsubsection{Gemini/FLAMINGOS-2 Spectrum}
\label{subsubsection:spectrum_gemini}

In addition, we included in our analysis a Gemini-South FLAMINGOS-2 spectrograph \citep{Eikenberry2004,Eikenberry2008} spectrum ($0.94-2.46\,{\rm \mu m}$) of COCONUTS-2b published by \citet{Zhang2025}. 
We downloaded the raw data together with the calibration files from the Gemini archive\footnote{\url{https://archive.gemini.edu/searchform}}, and we re-reduced the data using \texttt{PypeIt}\footnote{\url{https://pypeit.readthedocs.io/en/1.17.1/index.html}} \citep{pypeit:zenodo,pypeit:joss_pub,pypeit:joss_arXiv}. The spectrum was taken in two wavelength ranges: JH-band ($0.94-1.96\,{\rm \mu m}$), and K-band ($1.96-2.46\,{\rm \mu m}$). We decided to re-reduce only the JH-band, and include only the available photometry for the K-band (see Section~\ref{subsec:photometry}). 
Using the photometry, we estimated that the K-band contributes approximately $2\%$ of the bolometric flux. In addition, the SNR of the K-band is low (SNR\,$\approx 2$, \citealt{Zhang2025}). Therefore, we concluded that the photometry in the K-band provides a good approximation of the flux.
Following the description on \citet{Zhang2025}, we used the standard HIP 43762 to perform flux calibration, we used the ABBA pattern to remove the background contamination, and we performed the telluric correction using the standard star.
As a result we obtained a spectrum with an average resolution of ${\rm R}\,\sim\,900$ in the wavelength range $0.94-1.96\,{\rm \mu m}$, with SNR\,$\approx8$. We found that our reduction is consistent with the result from \citet{Zhang2025}. We included a detailed discussion on the comparison of the two spectra in Appendix~\ref{sec:appendix_comparison_red}.

\begin{deluxetable*}{ccccc}[ht!]
\tabletypesize{\scriptsize}
\tablecaption{Properties of COCONUTS-2. \label{table:coconuts2}}
\tablehead{ &  \colhead{COCONUTS-2A} & \colhead{Ref} &  \colhead{COCONUTS-2b} & \colhead{Ref}}
\startdata 
 & & Photometry & & \\\hline 
 GALEX/GALEX.FUV [mag] & ${17.704}\pm{0.161}$ & Bianchi2011 & - & -\\ 
GALEX/GALEX.NUV [mag] & ${16.459}\pm{0.065}$ & Bianchi2011 & - & -\\ 
SLOAN/SDSS.u [mag] & ${13.474}\pm{0.057}$ & Ahumada2022 & - & -\\ 
TYCHO/TYCHO.B [mag] & ${13.424}\pm{0.313}$ & Hog2000 & - & -\\ 
SLOAN/SDSS.g [mag] & ${12.155}\pm{0.067}$ & Ahumada2022 & - & -\\ 
$G_{\rm BP}$ (GAIA/GAIA3.Gbp) [mag] & ${11.558}\pm{0.055}$ & Gaia2023 & - & -\\ 
TYCHO/TYCHO.V [mag] & ${11.636}\pm{0.119}$ & Hog2000 & - & -\\ 
$G$ (GAIA/GAIA3.G) [mag] & ${10.162}\pm{0.054}$ & Gaia2023 & - & -\\ 
SLOAN/SDSS.r [mag] & ${10.541}\pm{0.078}$ & Ahumada2022 & - & -\\ 
SLOAN/SDSS.i [mag] & ${9.075}\pm{0.079}$ & Ahumada2022 & - & -\\ 
$G_{\rm RP}$ (GAIA/GAIA3.Grp) [mag] & ${8.995}\pm{0.055}$ & Gaia2023 & - & -\\ 
SLOAN/SDSS.z [mag] & ${8.600}\pm{0.090}$ & Ahumada2022 & - & -\\ 
Gemini/Flamingos2.Y [mag] & - & - & ${20.020}\pm{0.100}$ & Leggett2015\\ 
${\rm J_{2MASS}}$ (2MASS/2MASS.J) [mag] & ${7.417}\pm{0.057}$ & Skrutskie2006 & - & -\\ 
${\rm J_{MKO}}$ (NSFCam.J) [mag] & - & - & ${19.340}\pm{0.050}$ & Kirkpatrick2011\\ 
${\rm H_{MKO}}$ (NSFCam.H) [mag] & - & - & ${19.680}\pm{0.130}$ & Leggett2015\\ 
${\rm H_{2MASS}}$ (2MASS/2MASS.H) [mag] & ${6.835}\pm{0.061}$ & Skrutskie2006 & - & -\\ 
${\rm Ks_{2MASS}}$ (2MASS/2MASS.Ks) [mag] & ${6.563}\pm{0.058}$ & Skrutskie2006 & - & -\\ 
${\rm K_{MKO}}$ (NSFCam.K) [mag] & - & - & ${20.030}\pm{0.200}$ & Leggett2015\\ 
W1 (WISE.W1) [mag] & ${6.477}\pm{0.071}$ & Marocco2021 & ${17.080}\pm{0.036}$ & Marocco2021\\ 
$[3.6]$ (IRAC.I1) [mag] & - & - & ${16.432}\pm{0.036}$ & Kirkpatrick2011\\ 
$[4.5]$ (IRAC.I2) [mag] & - & - & ${14.621}\pm{0.020}$ & Kirkpatrick2011\\ 
W2 (WISE.W2) [mag] & ${6.287}\pm{0.058}$ & Marocco2021 & ${14.610}\pm{0.015}$ & Marocco2021\\ 
AKARI/IRC.S9W [mag] & ${6.036}\pm{0.069}$ & Ishihara2010 & - & -\\ 
MIRI.F1000W [mag] & - & - & ${12.910}\pm{0.005}$ & This work\\ 
IRAS/IRAS.12mu [mag] & ${6.147}\pm{0.121}$ & Abrahamyan2015 & - & -\\ 
W3 (WISE.W3) [mag] & ${6.311}\pm{0.056}$ & Cutri2021 & ${11.911}\pm{0.155}$ & Cutri2021\\ 
MIRI.F1280W [mag] & - & - & ${12.386}\pm{0.005}$ & This work\\ 
MIRI.F1800W [mag] & - & - & ${12.066}\pm{0.010}$ & This work\\ 
W4 (WISE/WISE.W4) [mag] & ${6.077}\pm{0.063}$ & Cutri2021 & - & -\\ 
\hline 
  & & Astrometry and Kinematics & & \\\hline 
 Epoch [yr] & $2016$ & Gaia2023 & $2025$ & This work\\ 
ra [deg] & $117.30085$ & Gaia2023 & $117.78671$ & This work\\ 
dec [deg] & $-76.70272$ & Gaia2023 & $-76.5804$ & This work\\ 
pmra [mas/yr] & ${-102.15}\pm{0.02}$ & Gaia2023 & ${-104.80}\pm{2.80}$ & This work\\ 
pmdec [mas/yr] & ${-192.92}\pm{0.02}$ & Gaia2023 & ${-189.70}\pm{4.50}$ & This work\\ 
parallax [mas] & ${91.83}\pm{0.02}$ & Gaia2023 & ${97.90}\pm{3.70}$ & This work\\ 
rv [km/s] & ${1.19}\pm{0.61}$ & Gaia2023 & ${0.10}\pm{4.50}$ & Faherty2025\\ 
\hline 
  & & Physical properties & & \\\hline 
 Age [Myr] & ${414}\pm{23}$ & This work & ${414}\pm{23}$ & This work\\ 
$\lbol$ [erg/s] & ${(7.31\pm0.12)\times 10^{31}}$ & This work & ${(2.73\pm0.22)\times 10^{27}}$ & This work\\ 
$\log \lbol/\lsun$ & ${-1.719}\pm{0.007}$ & This work & ${-6.15}\pm{0.03}$ & This work\\ 
Mass [$\Msun, \Mjup  $] & ${0.40}_{-0.02}^{+0.01}$ & This work & ${7.50}_{-0.40}^{+0.40}$ & This work\\ 
Radius [$\Rsun, \Rjup  $] & ${0.366}_{-0.014}^{+0.005}$ & This work & ${1.122}_{-0.004}^{+0.005}$ & This work\\ 
$\log$g [dex] & ${4.91}_{-0.01}^{+0.02}$ & This work & ${4.17}_{-0.02}^{+0.02}$ & This work\\ 
Teff [K] & ${3552}_{-32}^{+65}$ & This work & ${493}_{-9}^{+9}$ & This work\\ 
\enddata 
\tablerefs{Abrahamyan2015: \citet{Abrahamyan2015}, Ahumada2022: \citet{Ahumada2022}, Bianchi2011: \citet{Bianchi2011}, Cutri2021: \citet{Cutri2021}, Faherty2025: Faherty in prep., Gaia2023: \citet{Gaia2023}, Hog2000: \citet{Hog2000}, Ishihara2010: \citet{Ishihara2010}, Kirkpatrick2011: \citet{Kirkpatrick2011}, Kirkpatrick2021: \citet{Kirkpatrick2021}, Leggett2015: \citet{Leggett2015}, Marocco2021: \citet{Marocco2021}, Skrutskie2006: \citet{Skrutskie2006}.
}\end{deluxetable*}

\subsection{Photometry}
\label{subsec:photometry}

\subsubsection{COCONUTS-2A}

COCONUTS-2A is a bright nearby M dwarf, which was observed by several surveys, therefore we were able to compile photometry points covering almost the full spectral energy distribution (SED, $0.15-22.09\,{\rm \mu m}$) from the NASA/IPAC Infrared Science Archive (IRSA)\footnote{\url{https://irsa.ipac.caltech.edu/}}. We show the compiled data in Table~\ref{table:coconuts2}.

\subsubsection{COCONUTS-2b}

We compiled the existing photometry in the literature for COCONUTS-2b, which we show in Table~\ref{table:coconuts2}. 
In addition, as part of JWST Cycle 1 we obtained three photometry points using the Mid-Infrared Instrument (MIRI) in the bands: F1000W ($8.8-11.1\,{\rm \mu m}$), F1280W ($11.3 - 14.3\,{\rm \mu m}$) and F1800W ($16.0 - 20.3\,{\rm \mu m}$)\footnote{The specific observations analyzed can be accessed via \dataset[https://doi.org/10.17909/rxm9-qd05]{https://doi.org/10.17909/rxm9-qd05}.}. 
For our analysis, we used the data which was generated using the calibration software version 1.17.1, and the files were created on 2025-03-19. 
For each of the three bands we used the photometry which was calculated using a $70\%$ encircled energy circular aperture in the Vega system by the JWST pipeline \citep{Dicken2024}. These three points can be found in Table~\ref{table:coconuts2}.



Using the new JWST spectrum in Figure~\ref{fig:opacities}, we estimated synthetic photometry for COCONUTS-2b, and obtained W1: $17.60\pm0.12$\,mag, W2: $14.52\pm0.03$, [3.6]: $16.97\pm0.12$ and [4.5]: $14.59\pm0.03$. 
To account for the flux in the gap of the JWST spectrum in the wavelength range $[3.68,3.79]\,{\rm \mu m}$, we applied a simple linear interpolation. 
When we compared the synthetic photometry to the values in the literature (Table~\ref{table:coconuts2}), we found that while [4.5] and W2 are consistent, the synthetic values for [3.6] and W1 are $0.5$\,mag fainter than the literature values.
These differences are consistent with the results from \citet{Beiler2024}, that compared synthetic photometry from low-resolution JWST spectra to the literature value of a sample of 23 late-type T and Y dwarfs. They found a median scatter of $0.3$\,mag in W1, which they note is consistent with the preflight uncertainty goal of $\sim 10\%$, and a systematic difference in the [3.6] band  of around $0.3$\,mag. 
As \citet{Beiler2024} found these differences across a large population of objects, it suggests that is not related to variability. The same conclusion was reached by \citet{Luhman2024}, who also suggested that the differences could be due to errors in the filter response.

To study the variability hypothesis, we downloaded the photometry (and corresponding epoch) available on IRSA for COCONUTS-2b on the W1 and W2 bands from the catalogs WISE \citep{Wright2010}, AllWISE \citep{Cutri2013,Marocco2021} and NEOWISE \citep{Mainzer2011}. 
In total we collected photometry spanning more than $13$ years, with a cadence of six months. 
We did not find significant trends of long term variability for W1. Using the NEOWISE data ---given that is the one with more photometry points--- we estimated a reduced $\chi$-squared value of $1.59$ ($\nu=154$), when compared to the average photometry, which clearly shows no variability.
In the case of W2, we did not find variability either (reduced $\chi$-squared of $0.83$ ($\nu=275$) for the NEOWISE data), but in this case all the data agrees well with the synthetic photometry. 

\subsection{Astrometry}
\label{subsec:parallax}

Astrometry for COCONUTS-2A is provided by \textit{Gaia} \citep{GaiaDR3}, while astrometry for COCONUTS-2b is available from \citet{Kirkpatrick2021}. The \citet{Kirkpatrick2021} values are based on dedicated NTT/SofI observations \citep{Smart2013} covering a baseline of $\sim$5.3 years. Several new NTT/SofI observations have been acquired as part of the NPARSEC program \citep{Smart2013} since \citet{Kirkpatrick2021}, and with the addition of the JWST/MIRI images presented here we derived updated parallax and proper motion for COCONUT-2b with data now spanning a baseline of nearly 12 years. 

The observing strategy and data reduction for the NTT/SofI data followed the procedure described in detail in \citet{Smart2013}. We refined the astrometric calibration of the SofI data by finding \textit{Gaia} sources in the images and fitting a transformation between their NTT and \textit{Gaia} DR3 coordinates. 

First, we detected and measured the centroid for all sources in each image using \texttt{imcore}\footnote{\url{http://casu.ast.cam.ac.uk/surveys-projects/software-release/imcore}}. The x, y pixel coordinates measured by \texttt{imcore} are converted to $\alpha, \delta$ (ra, dec) using each image's WCS, and then matched to \textit{Gaia} DR3 using a 3\arcsec\ matching radius. We found between $\sim$60 and $\sim$80 \textit{Gaia} sources in each NTT image. The positions of matching \textit{Gaia} sources are propagated from the \textit{Gaia} epoch (2016.0) to the epoch of each observation using their measured parallax and proper motion. We then derived a transformation between our measured coordinates and the \textit{Gaia} coordinates by projecting both onto a tangent plane whose tangent point is defined by the CRVAL1 and CRVAL2 FITS header keywords. Finally, we fit for a six-parameter transformation, which accounts for offsets, rotation, skew, and scaling. The parameters of the transformation were determined using the IDL routine \texttt{mpfit} \citep{Markwardt2009}\footnote{\url{http://cow.physics.wisc.edu/~craigm/idl/fitting.html}}. The residuals of the fit are added in quadrature to the measurement errors to compute the final coordinate uncertainties. 

Astrometric distortion was accounted for by deriving a distortion map. To do that, we took the post-calibration astrometric residuals (i.e. $\Delta\alpha = \alpha_{Gaia} - \alpha_{NTT}$, $\Delta\delta = \delta_{Gaia} - \delta_{NTT}$) for all stars from all NTT images, and produced a map of residuals as a function of position on the detector. We binned this map using a N$\times$N grid, and in each bin we determined the median residuals. We corrected the position of our target in each image by adding the $\Delta\alpha$, $\Delta\delta$ values for the appropriate bin.

JWST/MIRI data was calibrated following the same procedure, with two important differences. First, because we have few reference stars, we could only fit for a three-parameters transformation, which accounts only for offset and rotation. Second, we did not need to derive a distortion map, since MIRI images are distortion-corrected by the imaging pipeline \citep[see e.g. Section 7 in][]{Dicken2024}.

We then combined the JWST and NTT measurements and fit for parallax and proper motion following the procedure described in \citet{Kirkpatrick2019} and \citet{Kirkpatrick2021}. The results are presented in Table~\ref{table:coconuts2}. The new values are consistent within $1\sigma$ with those presented in \citet{Kirkpatrick2021}.

\section{Age estimation}
\label{sec:age}

Being a binary, COCONUTS-2 presents a valuable opportunity to break the luminosity-age-temperature degeneracy of brown dwarfs. In this work we present full kinematics for both components of the binary. \citet{Marocco2024} noted the potential membership of COCONUTS-2 to the Corona of the Ursa Major (CUMA) moving group, which has the same age as the Ursa Major (UMA) group, the core of CUMA: $414\pm 23$\,Myr \citep{Jones2015}. In this section, we describe how we confirmed that membership using the 3D kinematics of both objects, rotation period, metallicity and C/O ratio.

As shown in Table~\ref{table:coconuts2}, COCONUTS-2A was observed by \textit{Gaia} DR3, which provided precise 3D kinematics and position. In the case of COCONUTS-2b, the radial velocity was estimated by Faherty et al. in prep---another work in the series of the JWST GO 2124 program---, and in this work we re-measured the astrometry using the MIRI photometry (see Section~\ref{subsec:parallax}). 
In short, for the calculation of radial velocity Faherty et al. in prep split the spectrum in windows that slightly overlap with each other, and in each window they built a forward model combining the Sonora Diamondback atmosphere models \citep{Morley2024}, a gaussian Line Spread Function (with its width determined by only one parameter), a Doppler shift (one parameter), and a linear blaze function (two parameters). Then they run an MCMC to sample the parameter space and pick the best-fitting model. After the fit, they checked if the residuals were representative of the expected errors. If they were not, they identified what error inflation would make the residuals follow a normal distribution with $1\sigma$ deviation ($\chi ^2 = 1$). Then they re-run the MCMC fit with this error inflation term in the spectrum. This extra step of inflating the uncertainties allows to get more representative errors when the model is a poor fit to the data. After the fit, they inspected the radial velocity measurements as a function of wavelength and checked that there were no significant trends. Finally, they averaged the measurements to obtain the final radial velocity and uncertainty.

Taking advantage of the 3D kinematics and 3D positions of both components of COCONUTS-2, we used BANYAN $\Sigma$ \citep{Gagne2018} to estimate the membership of the binary. 
BANYAN $\Sigma$ uses Bayesian inference to estimate the probability of an object belonging to a known association within $300$\,pc. This probability is calculated by comparing the 3D position and kinematics of the object to the position and velocity models of each group.
Using BANYAN $\Sigma$, we estimated a $99\%$ probability for each of the components of belonging to the CUMA group. 
Because two of the parameters of COCONUTS-2b---radial velocity and parallax---exhibit large uncertainties, we performed a statistical test of the membership probability. We simulated $1000$ stars with a random radial velocity measurement drawn a normal distribution defined by the value and uncertainty of the radial velocity measurement for COCONUTS-2b. The rest of the parameters were left unmodified. We used Banyan $\Sigma$ to calculate the probability of membership of each of these $1000$ stars, and obtained $99\%$ probability of belonging to CUMA for all the iterations. We repeated this processes but this time the random parameter was parallax, and obtained the same result, confirming the robustness of the result.

\begin{figure*}[ht!]
\begin{center}
\includegraphics[width=\linewidth]{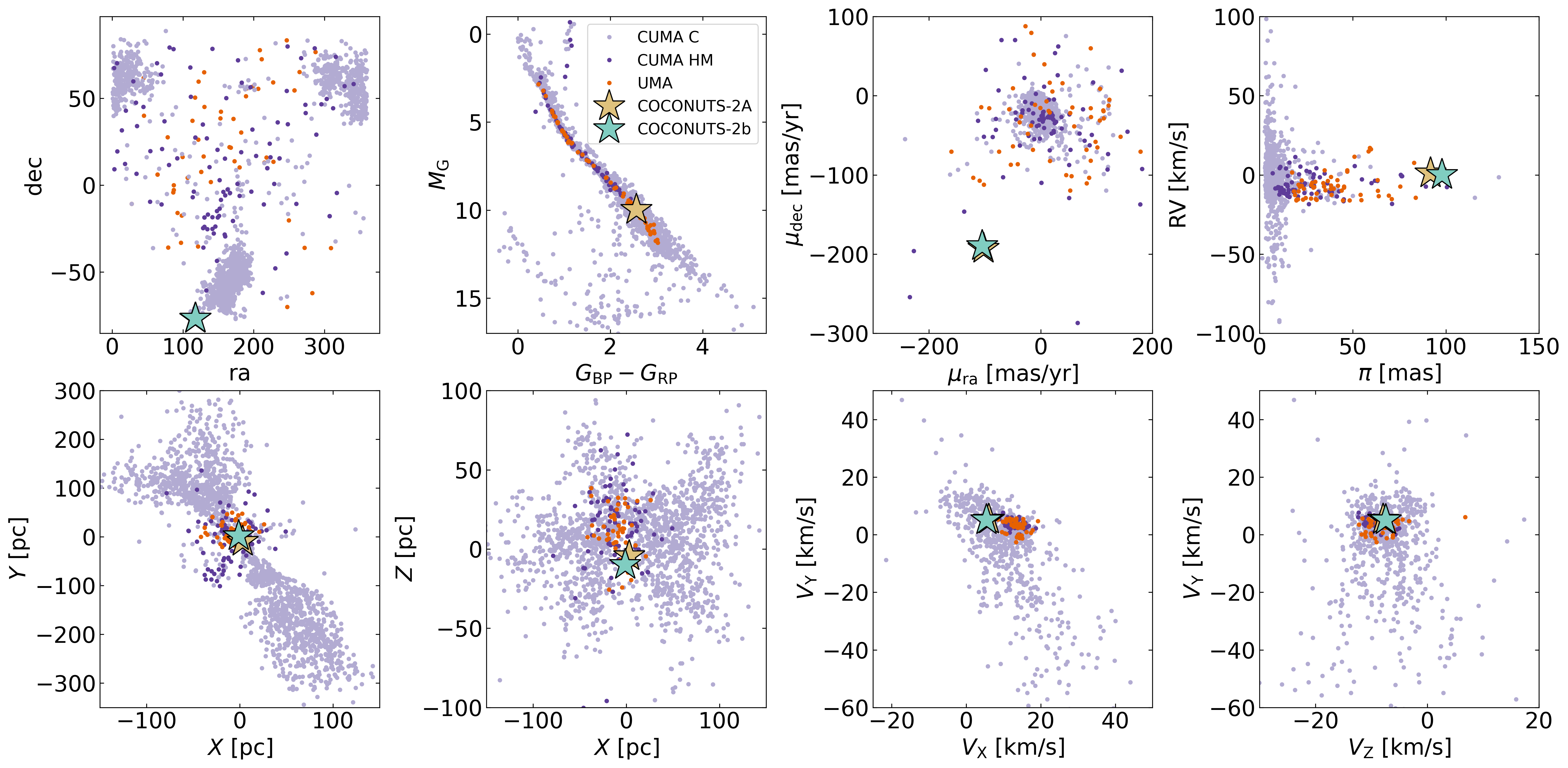}
\caption{Astrometric and photometric plots for the members of CUMA. The top row shows right ascension (ra), declination (dec), the \textit{Gaia} color-magnitude diagram, proper motion ra ($\mu _{\rm ra}$), proper motion dec ($\mu _{\rm dec}$), radial velocity (RV) and parallax ($\pi$). The bottom row shows the 3D galactic positions (X, Y and Z) and the 3D galactic velocities ($V_{\rm X}$, $V_{\rm Y}$ and $V_{\rm Z}$). We show in light-purple the candidate members (CUMA C) and in purple the high-likelihood members (CUMA HM). We obtained this classification from the MOCA database. We also include in orange the members of the UMA cluster from \citet{Capistrant2024}. In a yellow star we show COCONUTS-2A, and in green COCONUTS-2b. We find that the components of COCONUTS-2 agree well with the cluster in 3D velocity and position, while staying closer to each other, confirming that they are a binary.} 
\label{fig:comparison_CUMA}
\end{center}
\end{figure*}

To further confirm the membership of COCONUTS-2 to CUMA, we study the chance alignment probability, rotation period distribution, metallicity, and C/O ratio of the group. We describe each of these checks below.
We obtained the members of CUMA from the Montreal Open Clusters and Associations (MOCA) database\footnote{\url{https://mocadb.ca/}} (\citealt{Gagne2024}, Gagn\'e et al.\,in prep.). Besides the membership according to kinematics, the stars in MOCAdb are classified as confirmed, high-likelihood and candidate members of clusters or young associations, depending on if extra information is available, such as magnetic activity and rotation periods, which supports that they are young. 
In Figure~\ref{fig:comparison_CUMA} we show the $1782$ members of CUMA: $1669$ candidate members shown in light-purple and $113$ high-likelihood members in purple. We also included the $56$ targets identified as members of the Ursa Major (UMA) cluster---the core of CUMA---by \citet{Capistrant2024} in orange. We found that $26$ of these objects were also included in our list of CUMA members. The two star symbols of Figure~\ref{fig:comparison_CUMA} represent the two members of the COCONUTS-2 system. The 3D position and velocity of the two objects agree well with the members of the cluster. In addition, the two components have similar position and velocity, supporting that these two objects are in a binary, which agrees with previous results \citep{Zhang2021b}.

\subsection{Kinematics chance alignment} 
\label{subsec:chancealignment}

The members of the CUMA group have a large spread both in position and velocity, as can be seen in Figure~\ref{fig:comparison_CUMA}. Therefore chance association with the group is a possibility. 
Banyan $\Sigma$ includes a complex description of what a field star looks like, and compares it to models of the groups included. However, by default its priors are set up such that every association has a recovery rate of $\sim90\%$ when using a probability of $90\%$ threshold, with input uncertainties typical of \textit{Gaia}. There are documented probability-to-false-positive-rates mappings for each of the published association, but this is not the case of CUMA. Therefore we decided to do a Monte Carlo test as a first characterization of the expected false-positive rate in CUMA. To test how likely it is to identify a field star as a member of CUMA we selected $4000$ random stars from the $150$\,pc sample from \textit{Gaia} DR3 with radial velocities, and randomized the ra, dec, parallax and the uncertainties, but left the combination of pmra, pmdec and radial velocities untouched. For our test we replaced the radial velocities uncertainty, and tried seven cases: $1\%$, $10\%$, $20\%$, $30\%$, $40\%$, $50\%$ and $60\%$ uncertainty.
With a $50\%$ uncertainty in the radial velocity, which is close to the $45\%$ uncertainty of COCONUTS-2b (see Table~\ref{table:coconuts2}), we found that 1 in every $91$ stars got randomly assigned a membership to CUMA with a probability larger than $95\%$, which represents a $1.1\%$ chance of alignment ($1.7\%$ with a cut at $90\%$). We found that this probability decreases to $0.68\%$ for an uncertainty of $10\%$, and $0.65\%$ with $1\%$ uncertainty. This analysis shows that, even using only kinematics, the probability of a random assignment of membership to CUMA is small, but the contamination needs to be taken into account.

\subsection{Rotation periods} 

As an additional test to confirm the membership of COCONUTS-2 to the CUMA group, we used rotation periods. COCONUTS-2A has a measured rotation period of $2.83 \pm 0.28$ days \citep{Zhang2021b}, which we compared to the rest of the members of the cluster.
We measured rotation periods for all the members of the group using the data from the Transiting Exoplanet Survey Satellite \citep[TESS,][]{Ricker2015}.
We analyzed a sample of $690$ members of CUMA which are classified as candidates or high likelihood members, and which are brighter than $G<15$\,mag and have $(\gbp-\grp)>0.5$. 
We decided on this magnitude cut given that we found that $80\%$ of stars fainter than $15$\,mag exhibited flat light curves. 
For the light-curve extraction, we followed the procedure described in \citet{Popinchalk2023}, which we briefly summarize below.

We used $\texttt{Tesscut}$ \citep{Brasseur2019} to download TESS Full-Frame Image (FFI) cutouts of $40\times 40$ pixels size, and extracted the light curves from them.
We de-trended the light curves using two methods. The first is Causal Pixel Model (CPM), which we implemented using the package \texttt{unpopular} \citep{Hattori2022}. 
CPM is based on the idea that variations in brightness which are shared by several pixels are TESS systematics. The package \texttt{unpopular} models these variations by selecting neighboring stars around the target of interest and subtracting their combined signal from the target's light curve.
This method is particularly useful when the star is faint.
The second method we used is simple aperture photometry (SAP) around the target.
This method consists of extracting the light curve using a $4$-pixels circular aperture, which works best for bright stars where TESS systematics is less important \citep{Curtis2019b,Popinchalk2023}.
In order to apply these methods and inspect each light curve, we used the package $\texttt{Tess\_check}$ \citep{Curtis2019b}.

\begin{figure*}[ht!]
\begin{center}
\includegraphics[width=\linewidth]{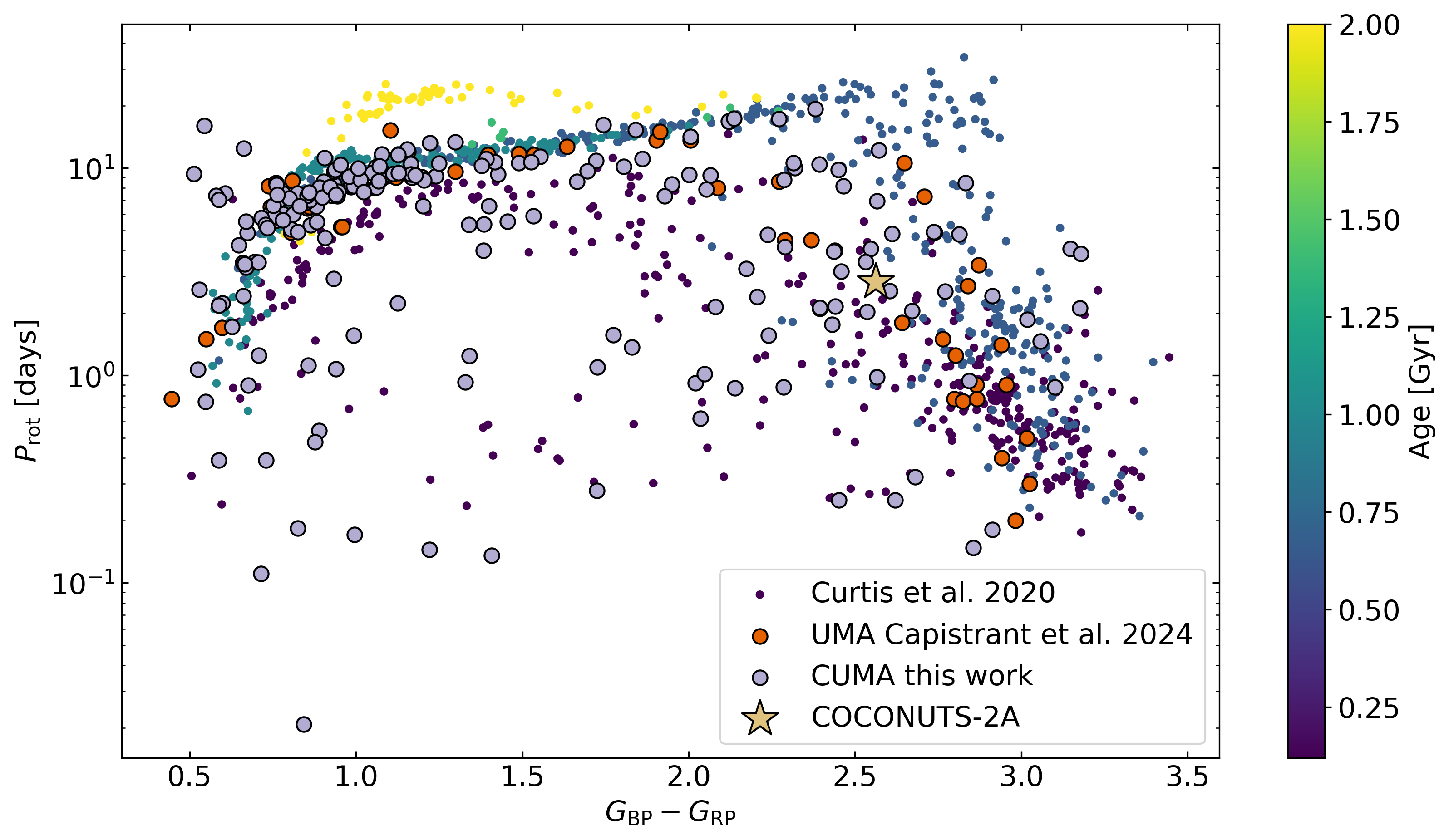}
\caption{Rotation period in days as a function of the \textit{Gaia} $(\gbp-\grp)$ color for CUMA. The candidates and high-likelihood members are shown in light-purple points. We included the members of the UMA cluster from \citet{Capistrant2024} in orange. In addition we show extra known young clusters with measured rotation periods, color-coded by age. We included the clusters compiled by \citet{Curtis2020}: Pleiades \citep[120 Myr,][]{Rebull2016}, Praesepe \citep[670 Myr,][]{Douglas2017, Douglas2019}, NGC 6811 \citep[1 Gyr,][]{Curtis2019b}, NGC 752 \citep[1.4 Gyr,][]{Agueros2018}, NGC 6819 \citep[2.5 Gyr,][]{Meibom2015} and Ruprecht 147 \citep[2.7 Gyr,][]{Curtis2020}. We also included COCONUTS-2A as a yellow five-point star, which has a measured rotation period. We found that the members of CUMA agree with the rotation period sequence of UMA, confirming that the two groups have the same age. We also found that the rotation period of COCONUTS-2A agrees with the locus of the CUMA group, providing further support for its membership in the group.} 
\label{fig:cuma_rotation}
\end{center}
\end{figure*}

After extracting the light curves, we used \texttt{LombScargle} from the package \texttt{astropy} to measure the rotation period of each star. 
The resulting rotation period measurements are in Table~\ref{table:cuma_clean} and are shown in Figure~\ref{fig:cuma_rotation} in light-purple circles. 
We removed stars with $\texttt{ruwe}>1.4$ from the figure, to discard possible binaries \citep{Fabricius2021}.
In addition, we included the rotation period measurements for the UMA cluster from \citet{Capistrant2024} in orange, and several known clusters color-coded by age \citep{Curtis2020}. The yellow five-point star in the figure represents the rotation period and color of COCONUTS-2A. We found that the sequence of rotation periods of CUMA agrees with the sequence of UMA, which supports that they have the same age. In addition, COCONUTS-2A agrees well with the locus of the CUMA group in the rotation versus color plot, supporting its membership to the group.

Using the rotation periods of the members of CUMA and UMA, we estimated the age of the two clusters with \texttt{ChronoFlow} \citep{Van-Lane2025}. \texttt{ChronoFlow} models the evolution of stellar rotation periods as a function of $(\gbp-\grp)$ color using a neural network-based probabilistic framework known as a conditional normalizing flow (CNF). This particular model is flexible to deal with complex density distributions, which was trained on a sample of known clusters with rotation period measurements. 
We obtained an age of $609^{+354}_{-276}$\,Myr for CUMA and $549^{+191}_{-230}$\,Myr for the UMA cluster, which further supports that both groups have the same age. In addition, the ages of both clusters are consistent with the previous age estimation for UMA ($414\pm23$\,Myr), which was calculated using stellar evolution models of a sample of A stars that belong to the cluster \citep{Jones2015}. 
We decided to continue using the previous age estimation given that the uncertainty is smaller. 

\begin{deluxetable}{cl}[!ht]
\tabletypesize{\scriptsize}
\tablecaption{Description of the columns in the sample of members of CUMA with rotation periods and metallicity measurements.}
\label{table:cuma_clean}
\tablehead{ \colhead{Column\tablenotemark{a}} & \colhead{Column description}} \ 
\startdata 
\texttt{moca\_oid}&ID internal to the MOCA database\\ 
\texttt{designation}&Normal designation of the target\\ 
\texttt{bp\_rp}&Gaia ($\gbp-\grp$) color\\ 
\texttt{period}&Rotation period (days)\\ 
\texttt{m\_h}&[M/H]\\ 
\texttt{m\_h\_error}&[M/H] uncertainty\\ 
\texttt{m\_h\_ref}&[M/H] reference\\ 
\texttt{fe\_h}&[Fe/H]\\ 
\texttt{fe\_h\_error}&[Fe/H] uncertainty\\ 
\texttt{fe\_h\_ref}&[Fe/H] reference\\ 
\texttt{c\_fe}&[C/Fe]\\ 
\texttt{c\_fe\_error}&[C/Fe] uncertainty\\ 
\texttt{c\_fe\_ref}&[C/Fe] reference\\ 
\texttt{o\_fe}&[O/Fe]\\ 
\texttt{o\_fe\_error}&[O/Fe] uncertainty\\ 
\texttt{o\_fe\_ref}&[O/Fe] reference\\ 
\texttt{ya\_prob}&Probability of belonging to a young association\\ 
\enddata
\tablenotetext{a}{Full table is available online.} 
\end{deluxetable}

\subsection{Metallicity} 
\label{subsec:metallicity}

As another method to test the membership of COCONUTS-2 to CUMA, we compared the metallicity of the primary and secondary to the members of the cluster. COCONUTS-2A has a measured metallicity of ${\rm [Fe/H]}=0.0\pm0.08$\,dex from \citet{Hojjatpanah2019}, and $-0.05 \pm 0.17$\,dex from \citet{Zhang2021b}. These measurements were made using high resolution spectra and the $(V-K)$-metallicity relation, respectively. We opted for the former, as it is more precise. COCONUTS-2b has an estimated metallicity from \citet{Zhang2025} who found sub-solar atmospheric [M/H] in the range $[-0.395,0.024]$\,dex, by fitting the Gemini/FLAMINGOS-2 spectrum with different atmospheric models. 
In this work, we performed a forward modeling analysis of both the Gemini/FLAMINGOS-2 spectrum and the JWST spectrum together, and we found that without any constrains on the fit, we obtained ${\rm [M/H]}=-0.337^{+0.016}_{-0.046}$\,dex, which is consistent with the results from \citet{Zhang2025}. For a detailed description of this analysis, see Section~\ref{subsection:forwardmodeling}.

\begin{figure}[ht!]
\begin{center}
\includegraphics[width=\linewidth]{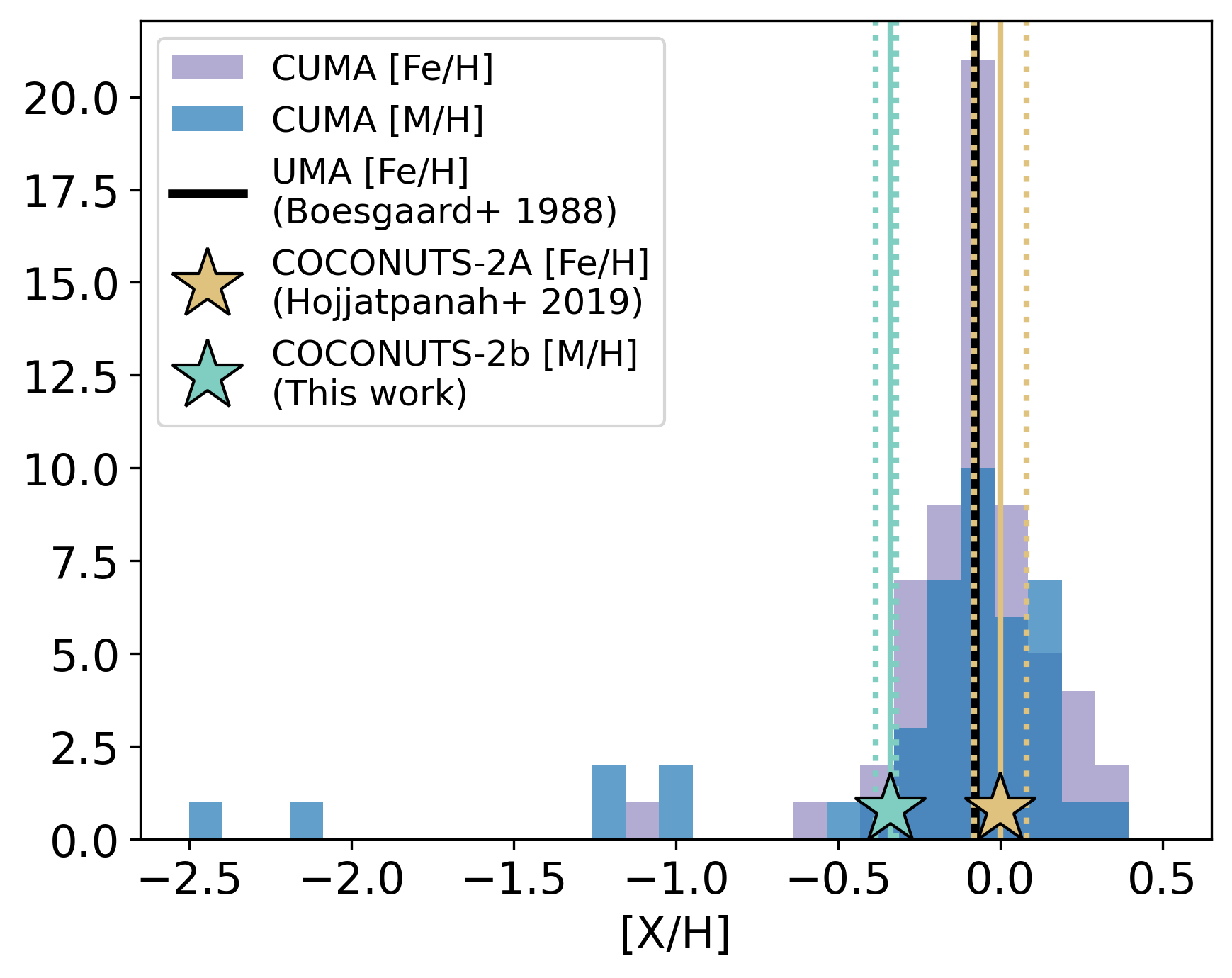}
\caption{[Fe/H] and [M/H] measurements for the candidate members of CUMA collected from the catalogs APOGEE-DR17, GALAH-DR3, RAVE-DR6 and LAMOST DR7. We included in a black horizontal line, the value of [Fe/H] for the UMA cluster \citep{Boesgaard1988}. We also included as a vertical line, the value of [Fe/H] for COCONUTS-2A and [M/H] for COCONUTS-2b (See Section~\ref{subsection:forwardmodeling}). The dotted vertical lines show the uncertainty in these two values. These measurements agree with the distribution of the moving group within uncertainty, supporting the membership of COCONUTS-2 to the group.} 
\label{fig:cuma_metallicity}
\end{center}
\end{figure}

To obtain metallicity measurements for the members of CUMA, we cross-matched the sample with APOGEE-DR17 \citep{Abdurrouf2022}, GALAH-DR3 \citep{Buder2021}, RAVE-DR6 \citep{Steinmetz2020} and LAMOST-DR7 \citep{Luo2019}. Comparisons of metallicity measurements for stars common to these catalogs indicate that they are consistent \citep{Carrera2019,Hegedus2023}. 
For each of the three catalogs we did a positional cross-match using a $3$\arcsec\ radius with the Tool for OPerations on Catalogues And Tables \citep[TOPCAT,][]{Taylor2005}. 
Applying the corresponding quality flags from each catalog to select the best abundance measurements, we obtained $62$ measurements of [Fe/H] and $46$ of [M/H]. 
The results for [Fe/H] and [M/H] together with the metallicity of COCONUTS-2A and 2b are shown in Figure~\ref{fig:cuma_metallicity}, and compiled in Table~\ref{table:cuma_clean}. 
We also included in a black line the median value of [Fe/H] for the UMA cluster measured by \citet{Boesgaard1988}: [Fe/H] = $-0.079\pm0.053$\,dex. This value was measured from high resolution spectra of the members of the group. 
The median value of [Fe/H] for the UMA cluster agrees well with the distribution of CUMA members. The uncertainty in the measurement from \citet{Boesgaard1988} corresponds to the dispersion in [Fe/H] values of the individual members, and shows a smaller scatter than CUMA. Below we discuss this point further.
The value of [Fe/H] of COCONUTS-2A is well centered in the distribution of the CUMA members, which supports the membership of the system to the group. 
The value of [M/H] of COCONUTS-2b is moderately sub-solar, however it agrees with the distribution of the cluster within uncertainty. One possible explanation for this difference is that there is a systematic uncertainty in the forward modeling analysis which has not been calibrated yet. We plan to explore this scenario in future studies.

\begin{figure}[ht!]
\begin{center}
\includegraphics[width=\linewidth]{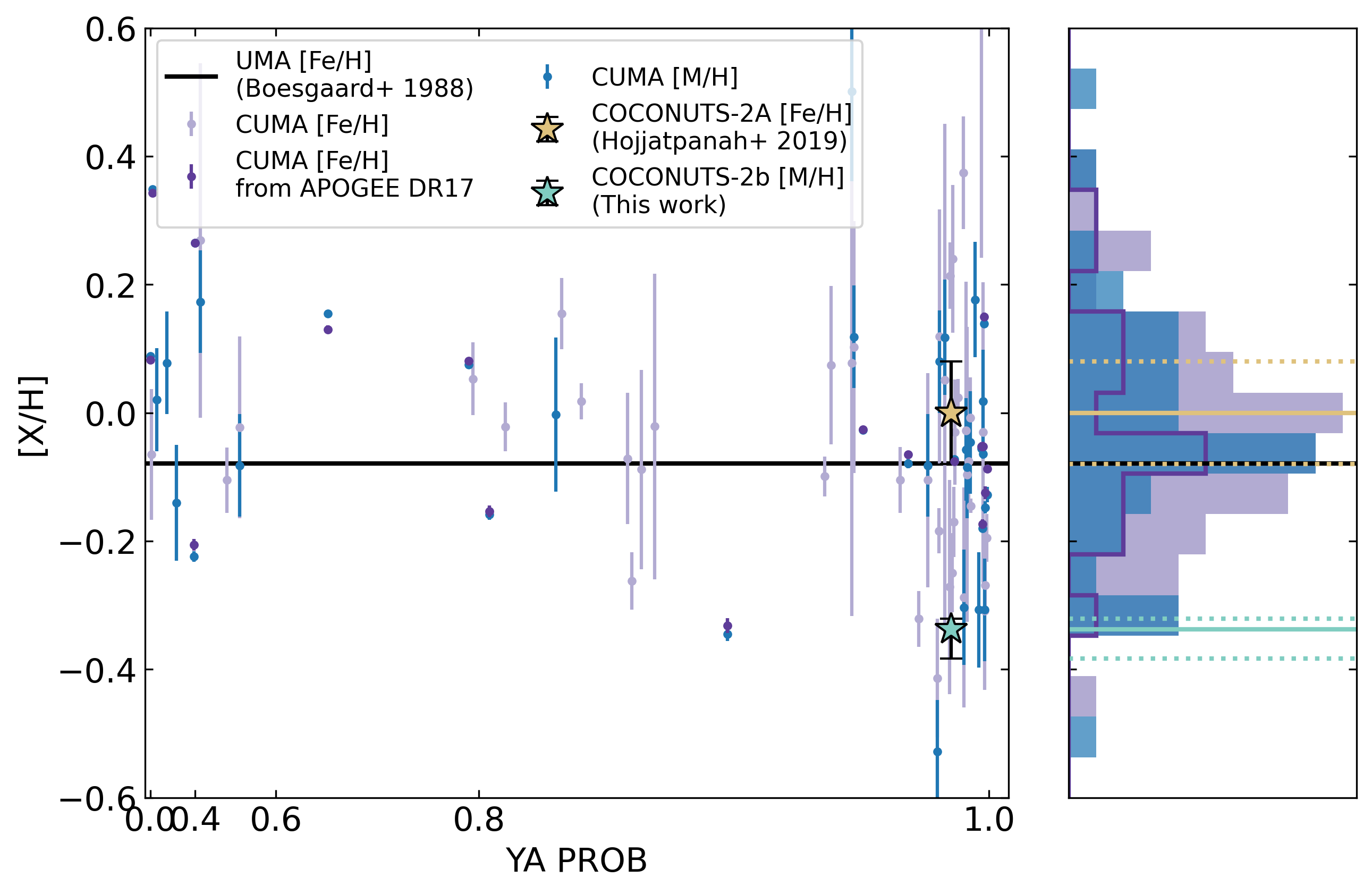}
\caption{[Fe/H] and [M/H] measurements for the candidate members of CUMA collected from the catalogs APOGEE-DR17, GALAH-DR3, RAVE-DR6 and LAMOST DR7. In the left panel we show the metallicity as a function of the probability of belonging to the CUMA group, and in the right panel is the distribution of values. We included the value of [Fe/H] for COCONUTS-2A and [M/H] for COCONUTS-2b as five-point stars in the left panel and as horizontal lines in the right panel. The dotted horizontal lines show the uncertainty in these two values. We also included in a black horizontal line, the value of [Fe/H] for the UMA cluster \citep{Boesgaard1988}.} 
\label{fig:cuma_metallicity_detail}
\end{center}
\end{figure}

We found $10$ candidate members of CUMA with metallicity significantly different from the rest of the group (8 with ${\rm [M/H]}<-0.5$ and two with ${\rm [Fe/H]}<-0.5$). We found that these outliers have a large uncertainty and/or are stars with lower probability of belonging to CUMA, meaning that they have the largest difference in position and velocity from the center of the cluster. Therefore, these objects are loosely bound or not members. 
In order to study the intrinsic scatter of the distribution of metallicities, we made Figure~\ref{fig:cuma_metallicity_detail}. This figure shows in the left panel the metallicity as a function of the probability of each candidate member to belong to the moving group, and in the right panel the histogram of the metallicity values.
Removing the extreme outliers, we estimated the median and the standard deviation of the [Fe/H] distribution for CUMA members, and we found $-0.06 \pm 0.25$\,dex, which agrees well with value for the UMA cluster with a significantly larger scatter. 
We found that the large scatter is in part due to the uncertainty in the measurements of metallicity from the different catalogs. 
We calculated the median uncertainty of the available measurements in our sample for each catalog, and found values of, in increasing order, APOGEE-DR17: $0.006$\,dex, LAMOST-DR7: $0.04$\,dex, GALAH-DR3: $0.06$ and RAVE-DR6: $0.2$\,dex. 
When using only stars from the APOGEE-DR17 catalog, the scatter in [Fe/H] is reduced to $0.17$\,dex, although there are only $18$ stars. Figure~\ref{fig:cuma_metallicity_detail} shows in dark purple the distribution of the metallicity for only APOGEE-DR17 stars, which clearly shows a smaller scatter for stars with high probability of belonging to the group. This figure also shows that most of the metallicity measurements are within $2\sigma$ from the value for UMA, and the stars with small uncertainty that differ from this value have lower probability of belonging to the CUMA moving group. Last, we calculated the reduced $\chi$-squared for the metallicity values of all the stars and only the stars with probability higher than $95\%$, and found $6.32$ ($\nu=61$) and $3.26$ ($\nu=38$), respectively. This shows that high-likelihood members agree better with a single metallicity population. In conclusion, the distribution of metallicities is consistent with a co-eval population.

\subsection{C/O ratio} 
\label{subsec:c_to_o}

\begin{figure}[ht!]
\begin{center}
\includegraphics[width=\linewidth]{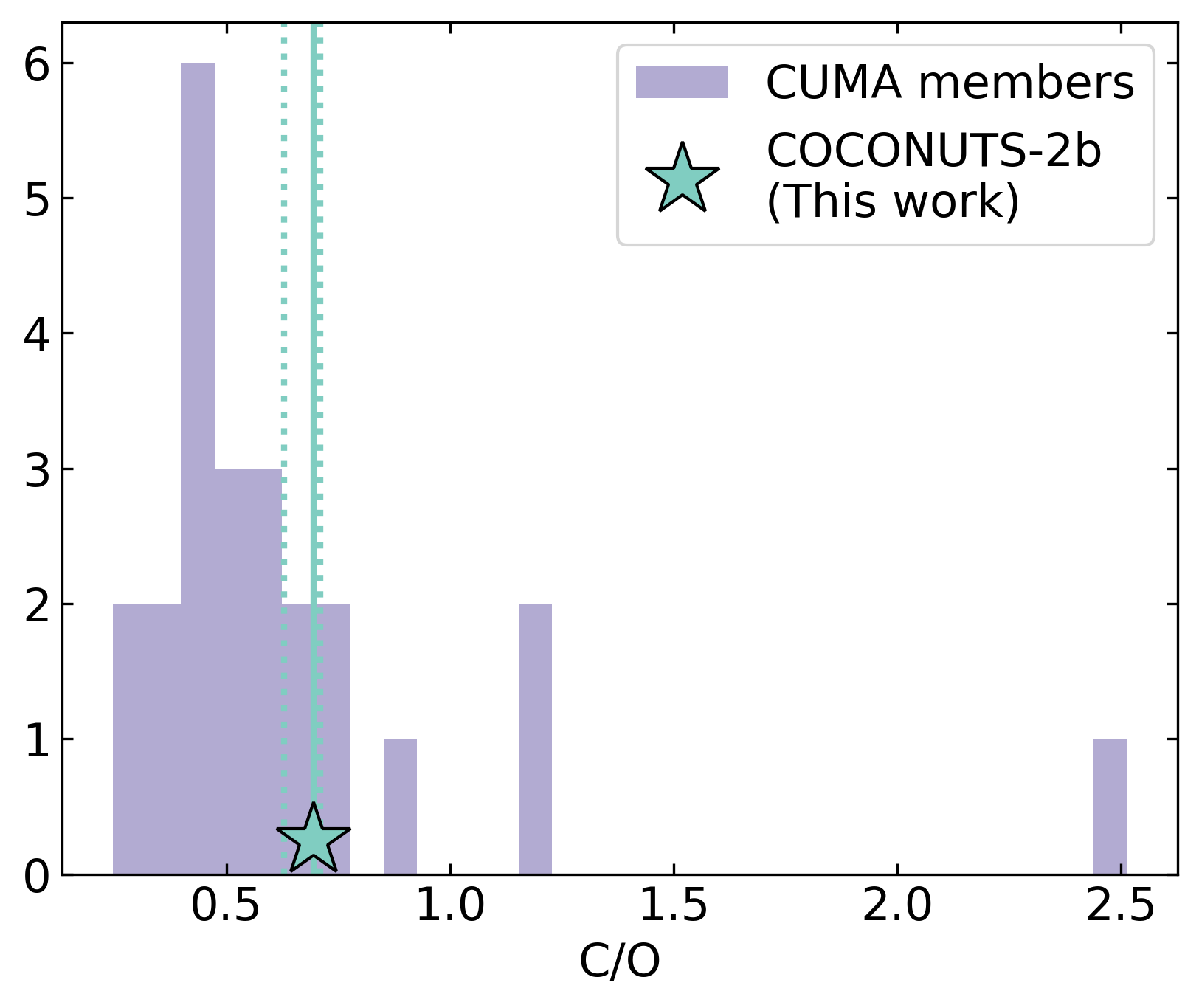}
\caption{C/O ratio measurements for the members of CUMA. These measurements were estimated from the $[\rm C/Fe]$ and $[\rm O/Fe]$ abundances from the catalogs APOGEE-DR17 and GALAH-DR3. We included in a vertical line the C/O ratio result for COCONUTS-2b from the forward modeling analysis in this paper (See Section~\ref{subsection:forwardmodeling}), which agrees with the values from the group. This suggests that COCONUTS-2b is a member of CUMA, and also it is likely to have formed in the same mechanism as a star.} 
\label{fig:cuma_c_to_o}
\end{center}
\end{figure}

As a final test of the membership of COCONUTS-2 to CUMA, we used the C/O ratio. This type of comparison has been applied to T~dwarfs in wide binaries with main-sequence stars \citep{Line2015,Phillips2024}, and planetary mass objects \citep{Hoch2023}. 
There is currently no measurement of the C/O ratio of COCONUTS-2A, therefore we leveraged the additional members of CUMA.
Taking advantage of the abundances compiled for the members of CUMA described above, we used $[\rm C/Fe]$ and $[\rm O/Fe]$, to estimate C/O. Using that the standard definition of abundances is

\begin{equation*}
    [{\rm X_1/X_2}] = \log_{10} \left( \frac{N_{\rm X_1}}{N_{\rm X_2}} \right)_{\rm star} - \log_{10} \left( \frac{N_{\rm X_1}}{N_{\rm X_2}} \right)_{\odot}
\end{equation*}

\noindent we derived that 

\begin{equation*}
    {\rm C/O} = \frac{N_{\rm C}}{N_{\rm O}} = \frac{10^{[\rm C/Fe] + [Fe/H]+\log_{10} \left( \frac{N_{\rm C}}{N_{\rm H}} \right)_{\odot}}}{10^{[\rm O/Fe]+[Fe/H]+\log_{10} \left( \frac{N_{\rm O}}{N_{\rm H}} \right)_{\odot}}}
\end{equation*}

\noindent where $\log_{10} ( N_{\rm C}/N_{\rm H})_{\odot}=8.43-12$, and $\log_{10} (N_{\rm O}/N_{\rm H})_{\odot}=8.69-12$ \citep{Asplund2009}.
Using this method we obtained $24$ measurements of C/O ratio for the sample.
We show the results from the estimation of C/O for the members of CUMA in Figure~\ref{fig:cuma_c_to_o}, and compiled in Table~\ref{table:cuma_clean}. 
Similar to the metallicity analysis in Section~\ref{subsec:metallicity}, we found that most of the values of C/O group around the solar value of $0.55 \pm 0.10$ \citep{Asplund2009}, with some outliers which have large uncertainties (SNR $< 2$).

COCONUTS-2b has a measured C/O ratio from \citet{Zhang2025}, that performed a forward modeling analysis of the Gemini/FLAMINGOS-2 spectrum, and found a result close to the solar value ($0.505^{+0.007}_{-0.004}$). 
In our work, we combined the Gemini/FLAMINGOS-2 spectrum with our JWST spectrum to compare to models. With the unconstrained fit, we found ${\rm C/O}=0.69^{+0.02}_{-0.07}$, which is consistent within $2\sigma$ with a solar value, and with the distribution of values from CUMA. We found that given the uncertainties in the JWST spectrum, our result is compatible with a solar value. We discuss this analysis further in Section~\ref{subsection:forwardmodeling}. 
We show our result for the C/O ratio together with the distribution of values for CUMA in Figure~\ref{fig:cuma_c_to_o}.
In addition, Lacy et al. in prep performed a similar analysis for COCONUTS-2b using both the Gemini/FLAMINGOS-2 and JWST spectra, but using a different forward modeling code, and found a solar C/O ratio (private communication).
Furthermore, there is an upcoming paper by Copeland et al. in prep that will present the results from the retrieval analysis for this object.

\begin{figure}[ht!]
\begin{center}
\includegraphics[width=\linewidth]{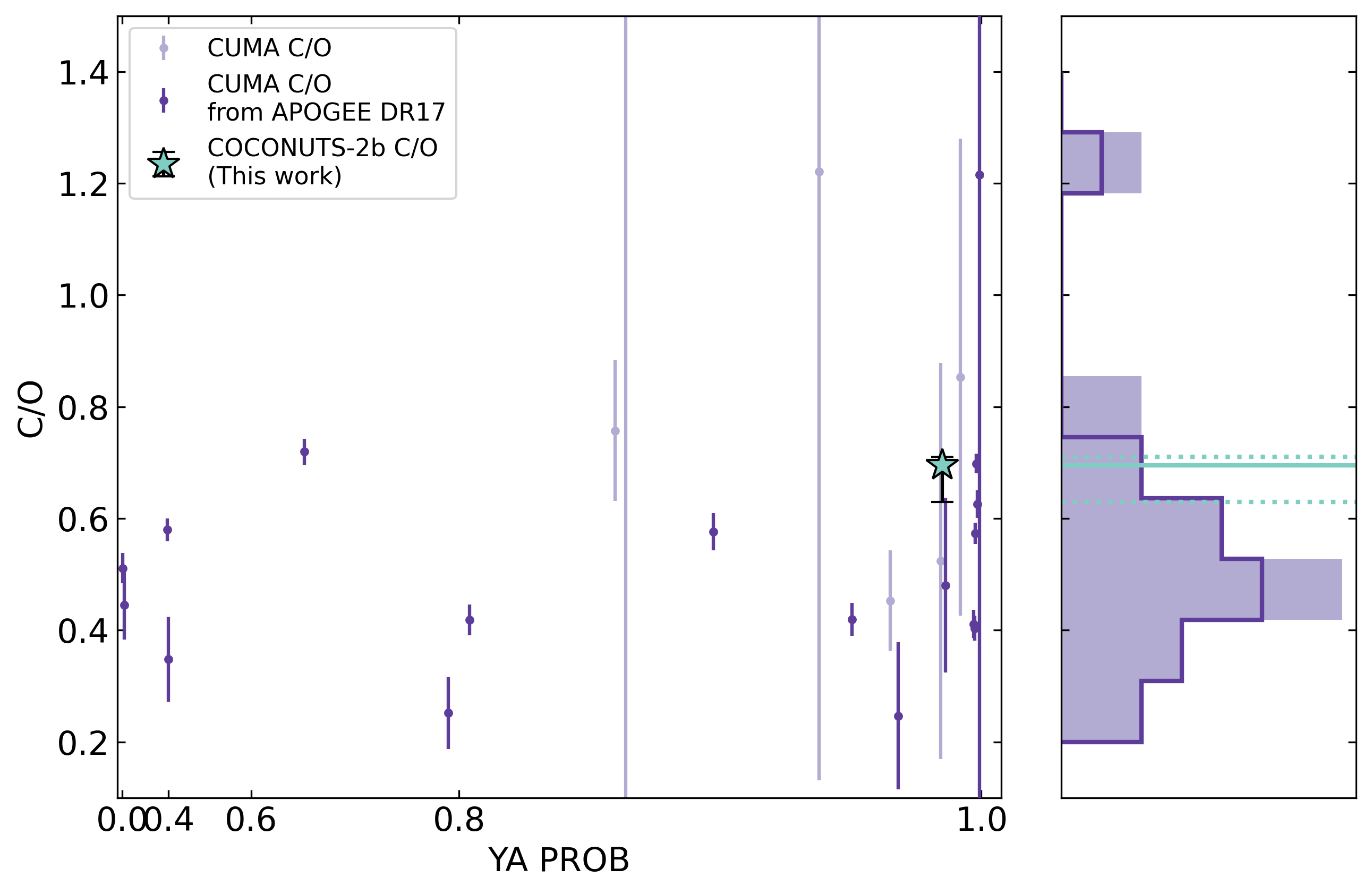}
\caption{We show in the left panel C/O ratio measurements for the members of CUMA as a function of the probability of each candidate to belong to the CUMA group, and in the right panel the histogram of the values. We also show as a five-point star in the left panel, and as a horizontal line in the right panel, our result for the C/O ratio for COCONUTS-2b from the forward modeling analysis.} 
\label{fig:cuma_c_to_o_detail}
\end{center}
\end{figure}

In order to study the intrinsic scatter of the C/O measurements, we made Figure~\ref{fig:cuma_c_to_o_detail}. We show in the left panel of the figure the values of C/O ratio as a function of the probability of belonging to the CUMA group, and in the right panel the distribution for the values. We included the value for COCONUTS-2b as a five-point star in the left panel and as a horizontal line in the right panel. This figure shows that the values that deviate from the solar C/O ratio have the largest uncertainty.
We also estimated the reduced $\chi$-squared value for the C/O values of all the stars and only the stars with probability higher than $95\%$, and found $1.34$ ($\nu=23$) and $1.05$ ($\nu=12$), respectively. This shows that the population is consistent with a constant C/O ratio value.

The agreement of the C/O ratio of COCONUTS-2b with the members of CUMA, supports its membership to the group. The C/O ratio also works as an indicator of the formation of the system. Planets that formed beyond the water snowline can have high values of C/O, even if the star has a solar C/O ratio \citep{Madhusudhan2011,Oberg2011}. Therefore, our results also suggest that COCONUTS-2b formed in the same mechanism as a star, supporting previous results \citep{Zhang2021b,Zhang2025}.

\section{Luminosity, Effective temperature, Mass, Radius and Surface gravity estimation}
\label{sec:lum}

Combining the spectra described in Section~\ref{subsec:spectra} and the photometry described in Section~\ref{subsec:photometry}, we used \texttt{SEDkit}\footnote{\url{https://github.com/BDNYC/sedkit/tree/main}} \citep{Filippazzo2015,Filippazzo2025} to estimate bolometric luminosity ($\lbol$), mass, radius, surface gravity ($\log g$) and effective temperature ($\teff$) for both components of COCONUTS-2. 
Below, we provide a brief description of how the code estimates each parameter, followed by detailed explanations for COCONUTS-2A and COCONUTS-2b, as well as the results.

\texttt{SEDkit} combines the input data, and extrapolates the missing flux at short and long wavelengths using Wien’s approximation and the Rayleigh–Jeans (RJ) law, respectively, using an initial guess for $\teff$, to compile a complete spectral energy distribution (SED). 
\texttt{SEDkit} then integrates this full SED to calculate the bolometric flux, and the bolometric luminosity using the parallax measurement. Combining $\lbol$ with the age estimated from CUMA (see Section~\ref{sec:age}) and a choice of evolutionary models, the code estimates mass, radius and $\log g$, and it estimates $\teff$ from the radius and luminosity using the Stefan-Boltzmann Law. We also estimated $\log \lbol/\lsun$ assuming a solar luminosity of \lbolsunnum. The uncertainty of the bolometric flux ($\fbol$) is determined using Monte Carlo propagation, accounting for flux and photometry uncertainties, assuming Gaussian distributions. Then the uncertainty of the $\lbol$ is propagated from errors in $\fbol$ and distance. Finally, the value and uncertainties for the rest of the parameters are estimated using a modification we added to \texttt{SEDkit}: the new code does a Monte Carlo propagation of uncertainties from normal distributions for the age and $\lbol$. Using the interpolation of the evolutionary models done with \texttt{LinearNDInterpolator} from \texttt{scipy}, it obtains a distribution for each of the other parameters.

\subsection{COCONUTS-2A} 

\begin{figure}[ht!]
\begin{center}
\includegraphics[width=\linewidth]{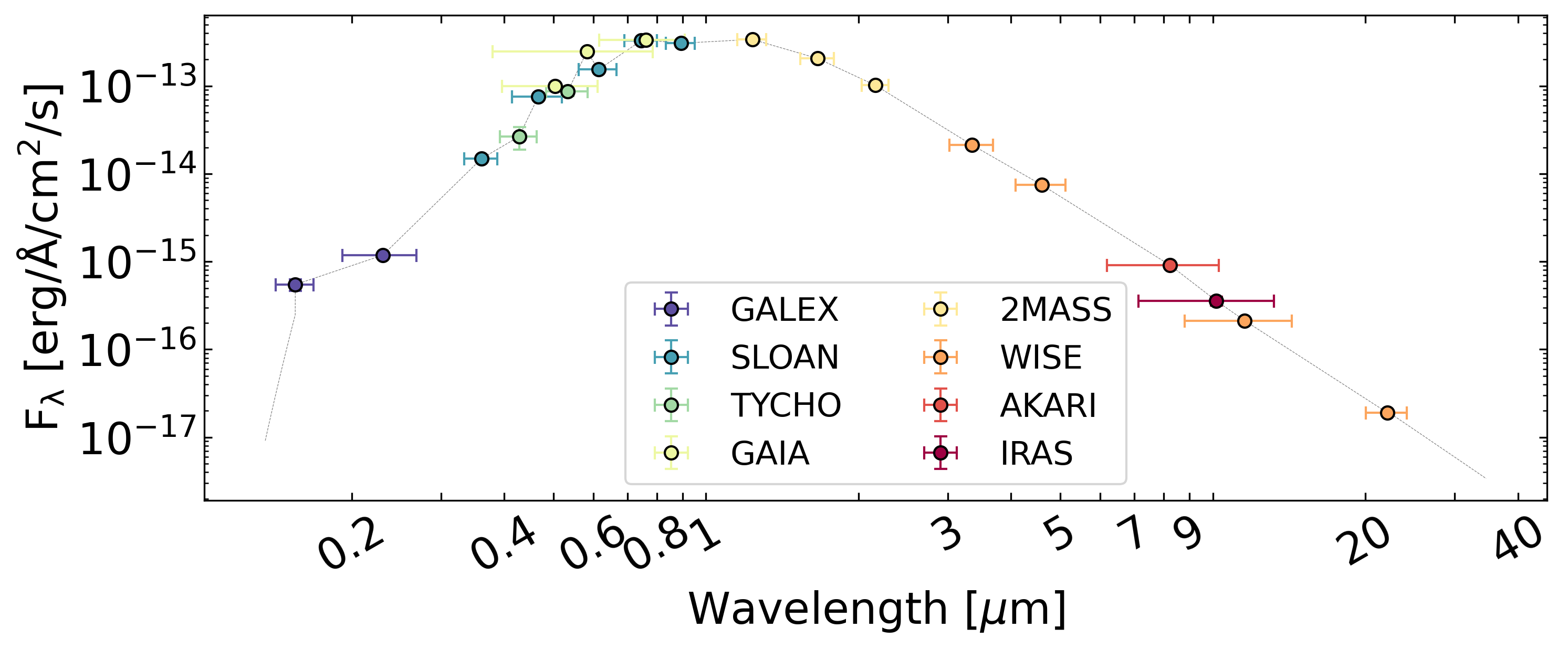}
\caption{Spectral energy distribution for COCONUTS-2A. We obtained the photometry from the literature (see Section~\ref{subsec:photometry}). The gray line represents the connecting lines, as well as the Wien and Rayleigh-Jeans approximations, extending beyond the existing data.} 
\label{fig:SED_coconuts2a}
\end{center}
\end{figure}

We assumed a $\teff=3406$\,K for the M3 component of the system for the Wien and RJ approximations, based on the value calculated by \citet{Gaidos2014}. 
To take into account model uncertainty in our parameter estimation, we used two different sets of evolutionary models: the MESA Isochrones \& Stellar Tracks \citep[MIST,][]{Choi2016,Dotter2016,Paxton2011,Paxton2013,Paxton2015,Paxton2018} and the \citet{Baraffe2015} models. 
The final value for each parameter is calculated as the average of the results from each model, with the uncertainty determined through error propagation. 
In both cases we used non-rotating, solar metallicity models. 
We show our results in Table~\ref{table:coconuts2}, and in Figure~\ref{fig:SED_coconuts2a} with a gray line showing the connecting lines used to interpolate the photometry points and calculate the bolometric luminosity. In addition, the gray line indicates the RJ and Wien approximations to extrapolate our data to low and high frequencies, respectively, where no data is available, as explained above in the short description of SEDkit.
Our results are consistent within $1\sigma$ with previous studies \citep{Gaidos2014,Zhang2021b} with an improvement of around $50\%$ in the uncertainties in most of the parameters due to our more complete SED, except for radius where they stay equal.

\subsection{COCONUTS-2b} 

\begin{figure*}[ht!]
\begin{center}
\includegraphics[width=\linewidth]{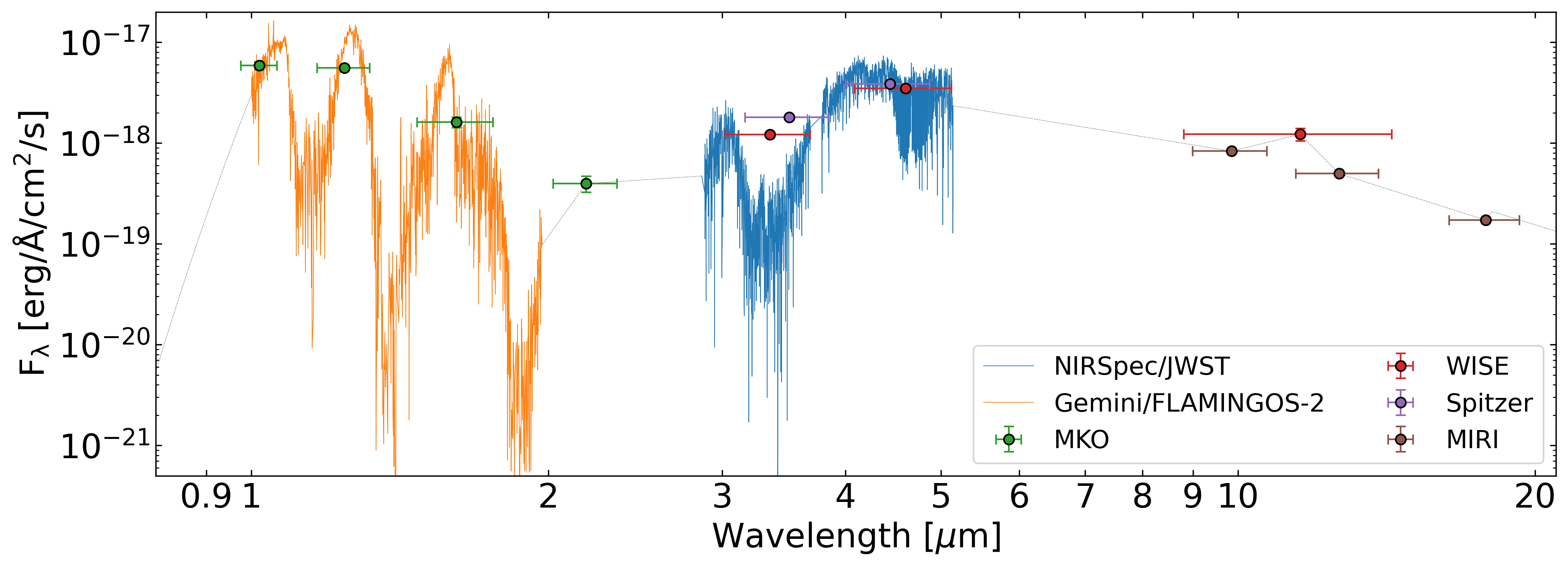}
\caption{Spectral energy distribution for COCONUTS-2b. We show in orange the Gemini/FLAMINGOS-2 spectrum \citep{Zhang2025}, in blue the G395H NIRSpec JWST spectrum, and in brown points the MIRI photometry. We also included photometry from the literature. MKO label (green, in increasing wavelength): Gemini/Flamingos2.Y, ${\rm J_{MKO}}$ (NSFCam.J), ${\rm H_{MKO}}$ (NSFCam.H) and ${\rm K_{MKO}}$ (NSFCam.K); WISE label (red): W1 (WISE.W1), W2 (WISE.W2), W3 (WISE.W3); Spitzer label (purple): [3.6] (IRAC.I1), [4.5] (IRAC.I2); MIRI label (brown): MIRI.F1000W, MIRI.F1280W, MIRI.F1800W. See Table~\ref{table:coconuts2} for the magnitude of each photometry point. The gray line represents the connecting lines, as well as the Wien and Rayleigh-Jeans approximations, extending beyond the existing data.} 
\label{fig:SED_coconuts2b}
\end{center}
\end{figure*}

We used an initial guess for $\teff$ of the T9 component of COCONUTS-2 of $483$\,K for the Wien and RJ approximations, based on the calculations from \citet{Zhang2025}. 
We also used the Sonora-Bobcat evolutionary models for solar metallicity \citep{Marley2021} to estimate the parameters of COCONUTS-2b using \texttt{SEDkit}. 
We show our results in Table~\ref{table:coconuts2}, and the compilation of data together with the approximations in Figure~\ref{fig:SED_coconuts2b}. The Gemini and JWST data account for $48\%$ of the bolometric flux, which increases to $68\%$ when including the three MIRI photometry points.
The parameters we estimated are within $1-2\sigma$ from previous studies \citep{Zhang2021b,Zhang2025}, but we found an $\lbol$ almost $4\%$ larger, which affected the rest of the parameters resulting in a higher mass and higher temperature. 

In Section~\ref{subsec:photometry} we discussed a systematic offset between the literature value of W1 and [3.6], and the synthetic photometry. We found that the synthetic photometry is $0.5$\,mag fainter than the literature value. Given that \citet{Zhang2021b,Zhang2025} used the literature photometry to estimate the $\lbol$ in this wavelength range, this would cause our luminosity to be fainter in comparison with their results. However, as we found the opposite effect, we conclude that the systematic difference is not strong enough to make a difference in the resulting values.

\section{Atmospheric composition}
\label{sec:atmosphere}

The strategically selected band of JWST high resolution spectra allows us to study COCONUTS-2b's atmospheric composition in detail, given that several dominant gaseous molecules absorb at these wavelengths, such as ${\rm CH_4}$, CO and ${\rm CO_2}$. 
In this section we present the JWST spectrum of COCONUTS-2b, compare it to different molecular opacities to identify the molecules present in the atmosphere, and perform a forward modeling analysis. 
Finally we compared our results to WISE J082507.35+280548.5, another object from the JWST GO 2124 program, which we find is a candidate member of CUMA.

\subsection{Molecules in the atmosphere of COCONUTS-2b}

We performed a detailed inspection of the JWST spectrum as shown in the top panel of Figure~\ref{fig:opacities}, where the resolution and precision of the JWST spectrum allowed us to interpret each of the features. 
In order to identify each molecule, we used opacities from The Data \& Analysis Center for Exoplanets (DACE)\footnote{\url{https://dace.unige.ch/opacity/}}, that we retrieved using the package \texttt{dace-query}\footnote{\url{https://dace-query.readthedocs.io/en/latest/index.html}}.
The DACE database contains opacities for different molecules which were calculated using the open-source code HELIOS-K \citep{Grimm2021}. 
HELIOS-K estimates the opacities using molecular lines from different databases. 
For our work we used opacities for CO, ${\rm CO_2}$ and ${\rm CH_4}$ which were calculated with the molecular lines from HITEMP \citep{Rothman2010,Hargreaves2020}.
In addition we included opacities for ${\rm H_2O}$ and ${\rm NH_3}$ which were calculated from the molecular lines included in the database ExoMol \citep{Barber2006,Yurchenko2011,Azzam2016,Tennyson2024}.
We show the opacities for these molecules in the bottom panel in Figure~\ref{fig:opacities}, assuming a temperature of $450$\,K and a pressure of $0.1$\,bar.
By comparing the opacities to the JWST spectrum, we see clear features of ${\rm NH_3}$, ${\rm CH_4}$, ${\rm CO_2}$, CO and ${\rm H_2O}$ which we marked approximately in the top panel Figure~\ref{fig:opacities}. 
We found clear narrow absorption features of ${\rm NH_3}$ in the range $[3.0,3.1]\,{\rm \mu m}$, and a broad absorption feature is generated by ${\rm CH_4}$ in the range $[3.1,4.15]\,{\rm \mu m}$, which is the dominant feature in the spectrum range we observed. In addition, there are clear wide absorption features from ${\rm CO_2}$ and CO in the ranges $[4.17,4.44]\,{\rm \mu m}$ and $[4.44,4.9]\,{\rm \mu m}$, respectively. There is clear overlap of the ${\rm CO}$ feature with the ${\rm H_2O}$ narrow absorption features in the range $[4.8,5.13]\,{\rm \mu m}$, and we further identify similar water signals in the range $[2.87,3.0]\,{\rm \mu m}$.

\subsection{Forward modeling}
\label{subsection:forwardmodeling}

We conducted a forward-modeling analysis of the data available for COCONUTS-2b, to study the atmospheric properties of this object. In this work, we fit the Gemini/FLAMINGOS-2 spectrum, analyzed by \citet{Zhang2025}, together with our JWST spectrum. For our analysis we used four atmospheric model grids: ATMO2020++ \citep{Leggett2021,Meisner2023}, PH3-free ATMO2020++ \citep{Leggett2024}, Sonora Elf Owl version 2 \citep{Mukherjee2024,mukherjee_2025_15150881}, and \citet{Lacy2023} models (LB23). 
We performed the forward modeling in a Bayesian framework using \texttt{SEDA} \citep{Suarez2021}\footnote{\url{https://seda.readthedocs.io/en/latest/}}, an open-source Python package for the forward modeling, and analysis of SEDs of ultracool objects and directly-imaged exoplanets. 
In this package, the sampling of the posterior is done using dynamic nested sampling \citep{Skilling2004,Skilling2006} which was implemented in the package \texttt{dynesty} \citep{Speagle2020,2024zndo..12537467K}\footnote{\url{https://dynesty.readthedocs.io/en/v2.1.5/}}. 
\texttt{SEDA} estimates the best fit from an interpolated grid of models without any prior information from evolutionary models. 
We refer to the results in Lacy et al. in prep, that performed a fit of COCONUTS-2b also combining the Gemini/FLAMINGOS-2 spectrum with the JWST spectrum, but assuming the Sonora-Bobcat evolutionary models \citep{Marley2021} to constrain the parameters. 

\subsubsection{Models included}

Here we briefly discuss the models selected for our analysis, and we refer to their individual publications for a more detailed description. We show a summary of all the models included in our analysis in Table~\ref{table:modelsincluded}. ATMO2020++ is based on the ATMO2020 framework \citep{Phillips2020}, which assumes radiative-convective equilibrium, solar metallicity, and cloud-free atmospheres. 
ATMO2020++ adds non-adiabatic thermal structure in the model grid, and includes metallicity as a grid parameter. Therefore it contains a grid of effective temperature, surface gravity and metallicity in the following ranges:
$T_{\rm eff}=[250,1200]$\,K, $\log g=[2.5,5.5]$\,dex, ${\rm [M/H]}=[-1,0.3]$\,dex. 
The ${\rm PH_3}$-free grid shares these assumptions, but excludes phosphine. Both ATMO2020++ and the ${\rm PH_3}$-free grid assume a solar C/O ratio, and diffusion coefficient of $K_{\rm zz}=10^{6}\,{\rm cm^2/s}$. The assumed model parameters are indicated in Table~\ref{table:modelsincluded} and in Table~\ref{table:forwardmodel} in parenthesis. 

The Sonora Elf Owl model grid builds on the Sonora-Bobcat model grid \citep{Marley2021}, which assumes radiative-convective equilibrium, rainout equilibrium chemistry, and cloud-free atmospheres, by incorporating disequilibrium chemistry, and an expanded range of [M/H] and C/O. The grid of the Sonora Elf Owl models covers a range of: $T_{\rm eff}=[275,2400]$\,K, $\log g=[3,5.5]$\,dex, $\log(K_{\rm zz})=[2,9]$\,dex, ${\rm [M/H]}=[-1,1]$\,dex and ${\rm C/O}\,=[0.5,2.5]$.

Last, the LB23 models assume radiative-convective equilibrium, and span cloud-free, cloudy, equilibrium, and disequilibrium conditions. The grid of the LB23 models covers the range: $T_{\rm eff}=[250,800]$\,K, $\log g=[3.5,5]$,
${\rm [M/H]}=[-0.5,0.5]$ and varying mixing lengths $H_{\rm mix}=[0.01,1]$, assuming a solar C/O ratio and $K_{\rm zz}=10^{6}\,{\rm cm^2/s}$. 

For each of the models described above, \texttt{SEDA} adds the radius of the object as a free parameter, which is estimated from the scaling factor required to match the model with the observed data, and the distance to the object. 

\begin{deluxetable}{cccccc}[!ht]
\tabletypesize{\scriptsize}
\tablecaption{Models included in the forward modeling analysis of COCONUTS-2b}
\label{table:modelsincluded}
\tablehead{ \colhead{Atmospheric} & \colhead{$\teff$} & \colhead{$\log g$} & \colhead{[M/H]}& \colhead{C/O}& \colhead{$\log(K_{\rm zz})$} \\ 
Model & [K] & [dex] & [dex] & & [${\rm cm}^{2}$/s] } \ 
\startdata 
\hline 
ATMO2020++&250:1200&2.5:5.5&-1:0.3&0.55&6.0\\ 
'' noPH3&250:1200&2.5:5.5&-1:0.3&0.55&6.0\\ 
Sonora Elf Owl&275:2400&3:5.5&-1:1&0.5:2.5&2:9\\ 
LB23&250:800&3.5:5&-0.5:0.5&0.537&6.0\\ 
\enddata
\end{deluxetable}

\subsubsection{Constrained versus unconstrained fits}

We performed the forward modeling analysis with two different sets of priors: 1) unconstrained with loose priors on the ranges for the parameters: $T_{\rm eff}=[400,700]$\,K, $\log g=[3.0,5.0]$\,dex and ${\rm [M/H]}=[-0.5,0.5]$\,dex; 2) constrained with fixed values for $T_{\rm eff}$ and $\log g$, where we selected the value from the grid of each model closest to the results from the SED fitting (See Section~\ref{sec:lum}). 
For both cases, we assumed a wide range of possible radius between $[0.7,1.2]\,\Rjup$. 
The fix values are indicated in Table~\ref{table:forwardmodel} in parenthesis. As \texttt{SEDA} does not assume evolutionary models to perform the fit, the constrained parameters allow us to identify the best fit assuming the age of the system ($414\pm23$\,Myr, see Section~\ref{sec:age}).
The results of the fit for both case scenarios are shown in Table~\ref{table:forwardmodel}, where we indicate the resulting reduced $\chi$-squared ($\chi ^2_\nu$) and the degrees of freedom ($\nu$) for each model. 
In both cases we found that the best fit model is the Sonora Elf Owl. This result differs from the results of \citet{Zhang2025}, that performed a forward modeling analysis of the Gemini/FLAMINGOS-2 spectrum, where the ${\rm PH_3}$-free ATMO2020++ models were preferred. 

For the unconstrained fit we found that the effective temperature is consistent within $2\sigma$ for most of the fits with the SED fit result ($T_{\rm eff}=493\pm9$\,K), while the surface gravity is much lower (SED result: $\log g=4.17\pm0.02$\,dex). 
This differs from the results from \citet{Zhang2025} that found surface gravities compatible with the SED results, but agrees with the results from Lacy et al. in prep. 
Given that the uncertainty of the JWST spectrum is significantly lower than the Gemini/FLAMINGOS-2 spectrum (See Section~\ref{subsec:spectra}), our fit is dominated by the JWST spectrum, indicating that the later is driving the difference in the results.
In addition, we found that for both the constrained and unconstrained fits we obtained radii close the 1\,$\Rjup$, expected for brown dwarfs \citep{Burrows1993,Burrows1997}, although our results are slightly smaller. This difference is probably due to the fact that in our fits the radius in a free parameters, instead of being constrained by evolutionary models.

For the metallicity measurement, we found that both in the constrained as in the unconstrained cases the value of [M/H] is sub-solar, with the unconstrained case resulting in a slightly more metal poor result. These results are consistent with the results in \citet{Zhang2025}. 

Lastly, we obtained a C/O ratio slightly higher than solar for the unconstrained case ($0.69^{+0.02}_{-0.07}$), and even higher for the constrained case ($0.87^{+0.08}_{-0.01}$). Both \citet{Zhang2025} and Lacy et al. in prep found a solar C/O ratio for COCONUTS-2b. To study this difference in detailed, we compared the JWST spectrum flux and uncertainty to different Sonora Elf Owl models, fixing all the parameters to the best fit indicated in Table~\ref{table:forwardmodel}, and varying only the C/O ratio. 
We found that given the uncertainty of the JWST spectrum, we cannot distinguish between the solar C/O ratio and the case of ${\rm C/O}=0.7$, but we can distinguish the case of ${\rm C/O}=0.8$. 
Therefore we concluded that for the unconstrained case, our results are compatible with solar C/O ratio, while for the constrained one they are not. 
This result is due to features such as ${\rm CH_4}$, CO and ${\rm CO_2}$ that are sensible both to C/O ratio, $T_{\rm eff}$ and $\log g$. Therefore, fixing the two latter parameters affects the result of the C/O ratio. 


\begin{deluxetable*}{lcccccccc}[!ht]
\tabletypesize{\scriptsize}
\tablecaption{Forward Modeling Results for COCONUTS-2b}
\label{table:forwardmodel}
\tablehead{ \colhead{Atmospheric Model} & \colhead{$\chi ^{2}_\nu$} & \colhead{$\nu$} &\colhead{$T_{\rm eff}$\tablenotemark{a}} & \colhead{$\log g$}& \colhead{[M/H]}& \colhead{C/O}& \colhead{$\log(K_{\rm zz})$}& \colhead{R} \\ 
  & & & [K] & [dex] & [dex] & & [${\rm cm}^{2}$/s] & [$\Rjup$]} \ 
\startdata 
&&&&Unconstrained&&&  \\ 
\hline 
 Sonora Elf Owl&4.24&4105&${516.61}_{-1.90}^{+5.83}$&${3.251}_{-0.087}^{+0.001}$&${-0.337}_{-0.046}^{+0.016}$&${0.69}_{-0.07}^{+0.02}$&${4.43}_{-0.13}^{+0.13}$&${0.846}_{-0.019}^{+0.011}$\\ 
LB23&8.47&4105&${500.00}_{-0.36}^{+0.34}$&${3.549}_{-0.031}^{+0.006}$&${-0.480}_{-0.056}^{+0.005}$&(0.537)&(6.0)&${0.818}_{-0.015}^{+0.002}$\\ 
ATMO2020++ noPH3&10.11&4107&${528.67}_{-4.91}^{+1.12}$&${4.010}_{-0.107}^{+0.006}$&${-0.332}_{-0.068}^{+0.008}$&(0.55)&(6.0)&${0.779}_{-0.005}^{+0.017}$\\ 
ATMO2020++&11.1&4107&${512.03}_{-1.17}^{+1.77}$&${3.736}_{-0.054}^{+0.009}$&${-0.492}_{-0.123}^{+0.006}$&(0.55)&(6.0)&${0.883}_{-0.009}^{+0.007}$\\ 
\hline 
 &&&&Constrained&&&  \\ 
\hline 
 Sonora Elf Owl&7.53&4105&(500.0)&(4.25)&${-0.179}_{-0.046}^{+0.004}$&${0.87}_{-0.01}^{+0.08}$&${4.03}_{-0.03}^{+0.42}$&${0.825}_{-0.003}^{+0.002}$\\ 
LB23&9.43&4105&(500.0)&(4.25)&${-0.239}_{-0.005}^{+0.008}$&(0.537)&(6.0)&${0.833}_{-0.002}^{+0.002}$\\ 
ATMO2020++ noPH3&11.14&4107&(500.0)&(4.0)&${-0.177}_{-0.004}^{+0.004}$&(0.55)&(6.0)&${0.931}_{-0.002}^{+0.002}$\\ 
ATMO2020++&11.58&4107&(500.0)&(4.0)&${-0.170}_{-0.004}^{+0.004}$&(0.55)&(6.0)&${0.972}_{-0.002}^{+0.002}$\\ 
\enddata
\tablenotetext{a}{Numbers in parenthesis indicate that the parameter is fixed in the model or the fit.} 
\end{deluxetable*}



\subsection{Comparison to a candidate member of CUMA: WISE J082507.35+280548.5}
\label{subsec:comparison0825}

In this section, we compare COCONUTS-2b to another object from the JWST GO 2124 program: WISE J082507.35+280548.5 (0825+2805, from now on), a Y0.5 dwarf discovered by \citet{Schneider2015}. 
By using BANYAN $\Sigma$ with the radial velocity measurement for 0825+2805 ($-18.7\pm3.4$\,km/s, Faherty et al. in prep, see Section~\ref{sec:age} for a description of the calculation) together with proper motion and parallax (pmra: $-66.7\pm0.9$\,mas/yr, pmdec:$-235.8\pm0.9$\,mas/yr, plx: $155.8\pm2.4$\,mas, \citealt{Kirkpatrick2021}), we calculated a probability of $93\%$ of belonging to CUMA. 
This percentage, although high, indicates a lower probability of belonging to CUMA than COCONUTS-2. 0825+2805 has a radial velocity measurement with a $20\%$ uncertainty, which given our calculations in Section~\ref{subsec:chancealignment}, indicates that the probability of chance alignment is $1$ in $89$. Therefore 0825+2805 is good candidate member of CUMA.

Rowland et al. in prep performed a retrieval analysis on the JWST spectrum of 0825+2805, and found that it has solar metallicity and solar C/O ratio (private communication). 
These measurements agree with the rest of the members of CUMA (see Section~\ref{sec:age}), which support the membership we calculated using kinematics.
Although more information is needed to confirm the membership of 0825+2805 to CUMA, the analysis done in this paper is enough to assume the membership for the rest of this section.
Therefore, 0825+2805 is a free floating object that, as candidate member of CUMA, is likely to have the same age as COCONUTS-2b. In addition, both objects have similar metallicity (with COCONUTS-2b being slightly sub-solar) and C/O ratio, making the comparison of the JWST spectra of the two objects interesting.

We combined the JWST/NIRSpec spectrum and the three MIRI photometry points obtained as part of the JWST GO 2124 program, with the data available in the literature which includes a spectrum from HST/WFC3 \citep{Schneider2015}, and photometry \citep{Kirkpatrick2021}, to measure the bolometric luminosity of 0825+2805 using \texttt{SEDkit}. 
Following the same method as described in Section~\ref{sec:lum}, we used an initial guess temperature of $350$\,K \citep{Leggett2024} to add the long and short wavelengths approximations, where data was not available. 
Assuming the age of CUMA, we estimated an effective temperature of $T_{\rm eff}=359\pm3$\,K, surface gravity of $\log g = 3.85\pm0.02$\,dex, mass of $M = {3.7}\pm{0.2}\,\Mjup$ and radius of $R = 1.144\pm0.002\,\Rjup$. 
Therefore, assuming the membership of 0825+2805 to CUMA, we found that it is a free-floating planetary mass object. 
This shows that CUMA is a good target for follow up using the Roman Space Telescope to study the population of free-floating planetary mass objects.

\begin{figure*}[ht!]
\begin{center}
\includegraphics[width=\linewidth]{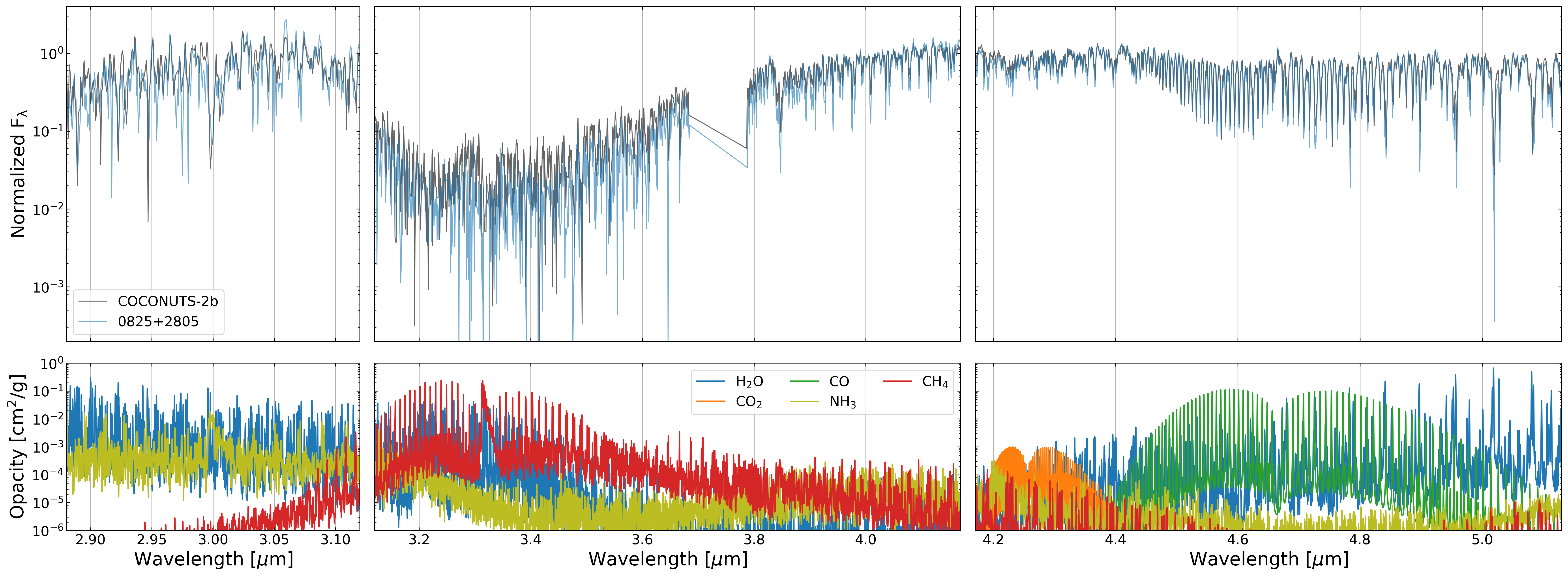}
\caption{Comparison of the G395H NIRSpec JWST spectra from COCONUTS-2b and 0825+2805 from the JWST GO 2124 program. In this figure, each of the three panels is normalized separately to facilitate the comparison to molecule abundances. Assuming that both objects have similar temperature-pressure profiles, we found that the abundance of ${\rm H_2O}$ and ${\rm CH_4}$ in the atmosphere of COCONUTS-2b are lower than for 0825+2805, which agrees with 0825+2805 being less massive and colder. We also found that the abundances of ${\rm CO}$ and ${\rm CO_2}$ are consistent, which is consistent with COCONUTS-2b being lower metallicity and it might indicates that the vertical mixing of 0825+2805 is stronger to compensate the effects of mass and temperature.} 
\label{fig:comparison_0825}
\end{center}
\end{figure*}

As mentioned above, comparing 0825+2805 and COCONUTS-2b presents a unique opportunity to isolate the effect of mass on brown dwarf evolution, since the two objects have similar ages and metallicity.
We show the JWST spectra of COCONUTS-2b and 0825+2805 in Figure~\ref{fig:comparison_0825}. 
We divided the spectra in three panels, and normalized each panel separately to facilitate the comparison of the different molecular features.
In addition, we included the opacities for the relevant molecules in the atmospheres of these two objects (${\rm H_2O}$, CO, ${\rm CO_2}$, ${\rm CH_4}$ and ${\rm NH_3}$).
We found that the feature of ${\rm H_2O}$ in the range $[4.8,5.13]\,{\rm \mu m}$ and ${\rm CH_4}$ in the range $[3.1,4.15]\,{\rm \mu m}$, are deeper for 0825+2805 than for COCONUTS-2b, which indicates that the atmosphere of 0825+2805 has a higher abundance of ${\rm H_2O}$ and ${\rm CH_4}$ than COCONUTS-2b. 
In addition, we found that the features of ${\rm CO_2}$ in the range $[4.17, 4.44]\,{\rm \mu m}$, and ${\rm CO}$ in the range $[4.44, 4.9]\,{\rm \mu m}$  have similar depths, which indicates that the abundances are similar in both objects.
For this analysis we are assuming that the temperature-pressure profiles of both objects are similar. A retrieval analysis of both objects is required to confirm that the depth of the feature can be directly related with the abundance.

We used the Sonora Elf Owl models to study the evolution of the molecular features discussed above with mass at a fixed age. We fixed the age to the age of CUMA, the metallicity and C/O ratio to the solar values, the $\log ({\rm K_{zz}})$ to $4.0$\,dex, and changed effective temperature and surface gravity with mass assuming the Sonora-Bobcat evolutionary models. 
We found that for more massive brown dwarfs (comparing from $4\,\Mjup$ to $10\,\Mjup$), effective temperature increases (from approximately $371$\,K to $591$\,K) and surface gravity increases (from approximately $4.06$\,dex to $4.3$\,dex), the ${\rm CH_4}$ and ${\rm H_2O}$ features shallow, while CO and ${\rm CO_2}$ deepen. 
Given that 0825+2805 is less massive and colder than COCONUTS-2b, this could explain that the abundances of ${\rm CH_4}$ and ${\rm H_2O}$ are higher for the former.

We also left all the parameters fixed to the values of COCONUTS-2b and changed only the metallicity from $-0.4$\,dex to solar metallicity, and found that while ${\rm CH_4}$ and ${\rm H_2O}$ do not change significantly, metal poor object have less CO and ${\rm CO_2}$ abundance than solar metallicity objects. Therefore the difference in metallicity between COCONUTS-2b and 0825+2805 might explain that the abundance of CO and ${\rm CO_2}$ are similar for both objects, given that it compensates the effect of mass and temperature.

Finally, we explored fixing all the parameters to the values of COCONUTS-2b, and changing only $\log ({\rm K_{zz}})$ from $4$\,dex to $8$\,dex. We found that CO and ${\rm CO_2}$ deepen with stronger mixing, while ${\rm CH_4}$ shallows, as it was pointed out by \citet{Mukherjee2024}. Therefore, this would indicate that the vertical mixing of 0825+2805 is stronger than that of COCONUTS-2b, given that it compensates the effect of mass and temperature in the abundances. This analysis shows the importance of benchmark brown dwarfs that belong to moving groups, which can be used to study atmospheric properties such as vertical mixing and metallicity.

\section{Conclusions}
\label{sec:conclusions}

In this work we study the COCONUTS-2 system, which is a wide binary, composed of an M3 star and a T9 brown dwarf. COCONUTS-2b is part of the JWST GO 2124 program, which obtained a NIRSpec spectrum with the G395H disperser, in the range $[2.87,5.13]\,{\rm \mu m}$, and three bands of MIRI photometry: F1000W ($8.8-11.1\,{\rm \mu m}$), F1280W ($11.3 - 14.3\,{\rm \mu m}$) and F1800W ($16.0 - 20.3\,{\rm \mu m}$). 
We used the new MIRI images to update the position, proper motions and parallax of COCONUTS-2b.
Taking advantage of the full kinematics of COCONUTS-2A from \textit{Gaia}, and our new astrometry together with the radial velocity obtained from the JWST spectrum for COCONUTS-2b (Faherty et al. in prep), we found that the COCONUTS-2 system is likely a member of the Corona of Ursa Major (CUMA) moving group, supporting previous results \citep{Marocco2024}. 
We further supported this membership by calculating the probability of chance alignment, measuring rotation periods for the group, and compiling metallicity and C/O ratio measurements for the members of CUMA. We found that the distributions are center at solar values for both the metallicity and C/O ratio.
By performing a forward modeling analysis of COCONUTS-2b, we found that is has a C/O ratio compatible with solar, and a slightly sub-solar metallicity.
By comparing the literature measurement of metallicity for COCONUTS-2A, and the results from the forward modeling for COCONTS-2b, we found good agreement between the system and the members of CUMA, supporting the membership. 
In addition, given that COCONUTS-2b has consistent C/O ratio with the rest of the members of CUMA, it likely formed with the same mechanism as a star, supporting previous results \citep{Zhang2021b,Zhang2025}. 

Using the new JWST/NIRSpec spectrum and MIRI photometry combined with a Gemini/FLAMINGOS-2 spectrum \citep{Zhang2025} and photometry from the literature \citep{Kirkpatrick2021}, we used \texttt{SEDkit} to estimate the bolometric luminosity, mass, effective temperature, surface gravity, and radius of COCONUTS-2b, assuming its age is that of the moving group. 
We found that COCONUTS-2b is a planetary mass object of $7.5\pm0.4\,\Mjup$.
In addition, by comparing the JWST spectrum with molecular opacities, we find clear evidence of presence of ${\rm NH_3}$, ${\rm CH_4}$, ${\rm CO_2}$, CO and ${\rm H_2O}$ in the atmosphere of COCONUTS-2b.

Finally we compared COCONUTS-2b to 0825+2805, a Y0.5 dwarf discovered by \citet{Schneider2015}, which we found is a potential member of CUMA ($93\%$ probability), and is also part of the JWST GO 2124 program. Furthermore, by doing a retrieval analysis of 0825+2805, Rowland et al. in prep found solar values for its metallicity and C/O ratio (private communication), which are similar to our forward modeling results for COCONUTS-2b, making these two objects interesting for comparison of the spectra. 
Assuming that the age of 0825+2805 is that of CUMA, we found that it is a cold ($359\pm3$\,K) planetary mass object ($3.7\pm0.2\,\Mjup$). 
By comparing the G395H NIRSpec JWST spectra for 0825+2805 with COCONUTS-2b, and assuming both objects have similar temperature-pressure profiles, we found that the former has higher abundances for ${\rm H_2O}$, ${\rm CH_4}$, which agrees with the fact that it is a colder and less massive object. In addition we found that both objects have ${\rm CO}$ and ${\rm CO_2}$ features with similar depths, which could be a result of COCONUTS-2b being slightly more metal poor, and/or could indicate that the vertical mixing is stronger for 0825+2805, to compensate the effects of temperature and mass.

\section{Acknowledgments}

RK would like to thank Peter R Eisenhardt and Jason Curtis for helpful discussions during the preparation of this manuscript.
Part of this work was carried out at the Jet Propulsion Laboratory, California Institute of Technology, under a contract with the National Aeronautics and Space Administration (80NM0018D0004). Program PID\#2124 is supported through contract JWST-GO-02124.002-A. 
This research made use of the Montreal Open Clusters and Associations (MOCA) database, operated at the Montr\'eal Plan\'etarium (\citealt{Gagne2024}, J. Gagn\'e et al., in prep.).
This publication makes use of The Data \& Analysis Center for Exoplanets (DACE), which is a facility based at the University of Geneva (CH) dedicated to extrasolar planets data visualisation, exchange and analysis. DACE is a platform of the Swiss National Centre of Competence in Research (NCCR) PlanetS, federating the Swiss expertise in Exoplanet research. The DACE platform is available at \url{https://dace.unige.ch}. EC and RAM acknowledge support from Chile FONDECYT/ANID \# 124 0049. RAM acknowledges support from Fondo GEMINI, Astrónomo de Soporte GEMINI-ANID grant \# 3223 AS0002.
JMV acknowledges support from a Royal Society - Research Ireland University Research Fellowship (URF/1/221932, RF/ERE/221108). 
BB acknowledges support from UK Research and Innovation Science and Technology Facilities Council [ST/X001091/1].
Some of the data presented in this paper were obtained from the Mikulski Archive for Space Telescopes (MAST) at the Space Telescope Science Institute. The specific observations analyzed can be accessed via \dataset[https://doi.org/10.17909/rxm9-qd05]{https://doi.org/10.17909/rxm9-qd05}. STScI is operated by the Association of Universities for Research in Astronomy, Inc., under NASA contract NAS5–26555. Support to MAST for these data is provided by the NASA Office of Space Science via grant NAG5–7584 and by other grants and contracts.

\begin{contribution}

RK performed most of the analysis for this work, coordinated the collaboration with co-authors and was responsible for writing and submitting the manuscript. 

CAB provided guidance and regular feedback during the development of the project, as well as funding for this project.

ARD together with RK were responsible for the forward modeling analysis and writing the section in the manuscript.

JKF provided regular feedback during the development of the project, and helped to coordinate the collaboration between co-authors. JKF is also the PI of the JWST GO program that obtained the spectrum and photometry for the objects studied in this work.

BL and GS provided valuable guidance with the forward modeling analysis, model comparison, and analysis of molecule abundances. Both co-authors contributed to refining the manuscript through detailed reviews.

JDK and FM estimated the new astrometry for COCONUTS-2b, and FM was responsible for writing Section~\ref{subsec:parallax}, describing this analysis.

JG provided valuable feedback on the membership analysis using Banyan $\Sigma$, and the tests done to estimate the chance alignment. 

JC and BB added significant discussion on the composition of the atmosphere of COCONUTS-2b, by comparing the results from this work with the retrieval analysis they are doing. 

NW helped with the forward modeling analysis by mentoring ARD at the beginning of the project. 

MJR improved significantly the comparison to the 0825+2805 object by sharing the results from the retrieval analysis, and discussing the results from the forward modeling of COCONUTS-2b.

DCBG, JMV, ACS and ECG improved the manuscript significantly with a thorough review of the text, and provided insightful comments which improved the work.

SAM provided valuable feedback on the estimation of bolometric luminosity of the object and development of \texttt{SEDkit}.

AR improved the manuscript significantly with a thorough review of the text.

RS, EC, RAM were responsible for obtaining the data we used in this work to estimate the astrometry of COCONUTS-2b.


\end{contribution}

\software{
           \texttt{astropy} \citep{astropy2013,astropy2018,astropy2022}; 
           \texttt{ChronoFlow} \citep{Van-Lane2025}; 
           \texttt{dace-query}; 
           \texttt{dynesty} \citep{Speagle2020,2024zndo..12537467K};
           \texttt{emcee} \citep{Foreman-Mackey2013}; 
           \texttt{imcore}; 
           \texttt{matplotlib} \citep{Hunter2007}; 
           \texttt{mpfit} \citep{Markwardt2009}; 
           \texttt{numpy} \citep{harris2020array}; 
           \texttt{PypeIt} \citep{pypeit:joss_arXiv,pypeit:joss_pub,pypeit:zenodo}; 
           \texttt{SEDA} \citep{Suarez2021}; 
           \texttt{SEDkit} \citep{Filippazzo2015,Filippazzo2025}; 
           \texttt{scipy} \citep{SciPy-NMeth2020}; 
           }

\appendix

\section{Comparison with the previous Gemini/FLAMINGOS-2 reduction}
\label{sec:appendix_comparison_red}

\begin{figure}[ht!]
\begin{center}
\includegraphics[width=0.6\linewidth]{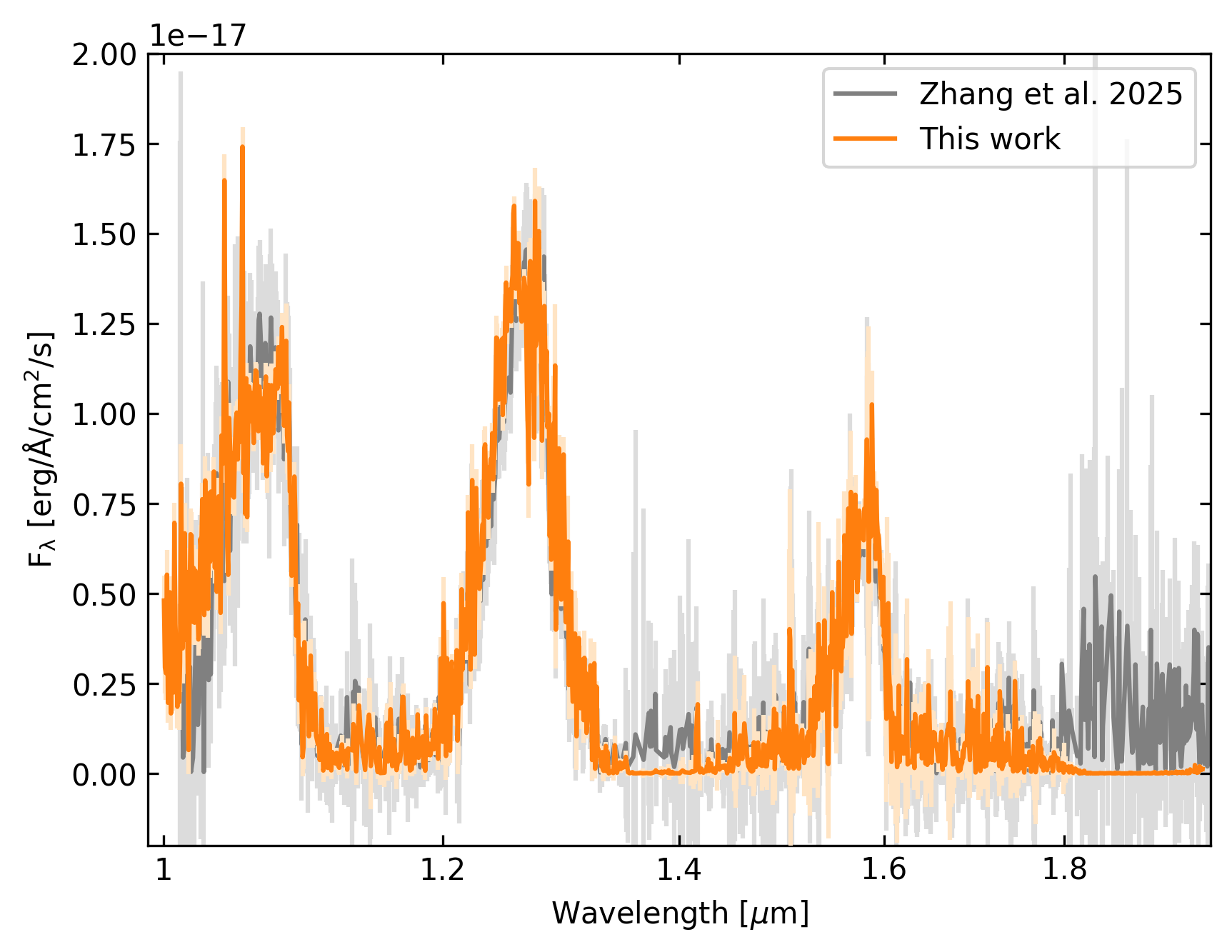}
\caption{Comparison of our reduction of the Gemini/FLAMINGOS-2 data with the original from \citet{Zhang2025}.} 
\label{fig:comparison_red}
\end{center}
\end{figure}

We compared the reduction of the Gemini/FLAMINGOS-2 data done by \citet{Zhang2025}, available online\footnote{\url{https://zenodo.org/records/13975825}}, with our reduction done with \texttt{PypeIt}, and the results are shown in Figure~\ref{fig:comparison_red}. We found in general good agreement between the two reductions. The areas where the two spectra differ most (around $1$\,$\mu$m, $1.4$\,$\mu$m and $1.9$\,$\mu$m) correspond to wavelength ranges where the telluric contamination is high, and the uncertainty of the flux is high. The spectra are still consistent within uncertainty. As the fit to the data was done using the uncertainty in the flux as a weight in both studies, these differences should not affect the results.

\pagebreak

\bibliographystyle{aasjournal}
\bibliography{bibliography,packages}

\end{document}